# The Best Time for an Update: Risk-Sensitive Minimization of Age-Based Metrics

Wanja de Sombre, Andrea Ortiz, Frank Aurzada, Anja Klein

***Abstract*—Popular methods to quantify transmitted data quality are the Age of Information (AoI), the Query Age of Information (QAoI), and the Age of Incorrect Information (AoII). We consider these metrics in a point-to-point wireless communication system, where the transmitter monitors a process and sends status updates to a receiver. The challenge is to decide on the best time for an update, balancing the transmission energy and the age-based metric at the receiver. Due to the inherent risk of high age-based metric values causing complications such as unstable system states, we introduce the new concept of risky states to denote states with high age-based metric. We use this new notion of risky states to *quantify* and *minimize* this risk of experiencing high age-based metrics by directly deriving the frequency of risky states as a novel risk-metric. Building on this foundation, we introduce two risk-sensitive strategies for AoI, QAoI and AoII. The first strategy uses system knowledge, i.e., channel quality and packet arrival probability, to find an optimal strategy that transmits when the age-based metric exceeds a tunable threshold. A lower threshold leads to higher risk-sensitivity. The second strategy uses an enhanced Q-learning approach and balances the age-based metric, the transmission energy, and the frequency of risky states without requiring knowledge about the system. Numerical results affirm our risk-sensitive strategies' high effectiveness.**

## I. Introduction

Exploiting Internet of Things (IoT) networks enhances monitoring in areas such as robotics and vehicular communication. This approach improves the resilience and robustness of applications [1]. Monitoring requires that IoT sensors transmit status updates regarding the monitored processes in a timely manner over unreliable, in general, wireless communication channels. The key challenge we investigate for monitoring scenarios is to identify the best times to send status updates. The strategy employed to determine these points in time must balance update freshness with the energy needed for transmission. We specifically examine such transmission strategies in a general monitoring scenario, divided into discrete time steps and consisting of a monitored process and a sensor transmitting status updates to a receiver.

To assess the freshness of the received status updates, the notion of Age of Information (AoI) was introduced by Kaul et al. [2]. AoI refers to the elapsed time since the generation of a status update and serves as a metric for determining an update's timeliness. The specific AoI requirements for status updates vary depending on the application. For instance, applications related to human safety often impose strict AoI requirements [3], which not only encompass a low average AoI, but also the minimization of the probability, or risk, of encountering large AoI values. Requirements like these necessitate the use of new, risk-sensitive transmission strategies.

Even though the definition of AoI is broad, there are several applications in which other age-based metrics are more appropriate. For example, when the significance of AoI is confined to specific time steps, a pull-based communication model is considered and a modified age-based metric called Query Age of Information (QAoI) is introduced [4]. The QAoI is defined by measuring the AoI only at designated *query time steps* while ignoring it at all other times. In other applications, updates are sensed at every time step, and the content of a packet has a decisive impact on when it should be sent. In other words, as long as the correct information about the monitored process is available at the receiver, the age of information is not important, but as soon as this information is incorrect, the age of information becomes relevant. Such cases have motivated the introduction of Age of Incorrect Information (AoII) [5]. The AoII is defined as zero every time the information at the sender and at the receiver are the same. Otherwise, and similar to the AoI, the AoII is increased by one for every time step, in which the information is different. Even though there are more variations of the AoI, e.g., Value of Information [6] or Age of Loop [7], their definition is dependent on the considered application. Therefore, we restrict ourselves to the AoI, QAoI, and AoII, due to their simplicity and broad applicability. We call these three metrics *age-based metrics*.

Research work on age-based metrics in wireless communication systems and, in particular, point-to-point wireless communication systems, have mainly focused on minimizing the average AoI [8]–[15]. In [8], queuing theory is used to derive closed-form expressions for the average AoI at the receiver under different queue models. In [9], the authors consider a monitoring system. They propose a strategy for the sender to

Wanja de Sombre, Andrea Ortiz and Anja Klein are with the Communications Engineering Lab, Technical University of Darmstadt, Germany ({w.sombre, a.ortiz, a.klein}@nt.tu-darmstadt.de).
Frank Aurzada is with the Probability and Statistics Group, Mathematics Departement, Technical University of Darmstadt, Germany (aurzada@mathematik.tu-darmstadt.de).
This work has been funded by the German Research Foundation (DFG) as a part of the projects B3 and C1 within the Collaborative Research Center (CRC) 1053 - MAKI (Project-ID 210487104) and has been supported by the BMBF project Open6GHub (Nr. 16KISK014) and the LOEWE Center EmergenCity. Part of this work was presented at the IEEE International Conference on Communications (ICC), Rome, Italy, June 2023 and at the IEEE Global Communications Conference (Globecom), Kuala Lumpur, Malaysia, 2023.

decide when to sample the monitored process and when to transmit a status update to the receiver. In a similar scenario, and assuming the sender monitors a dynamic Markov process, the authors in [10] exploit differential encoding to increase the system's reliability against transmission errors. In [11], a capacity-constrained point-to-point scenario is considered. Assuming that the transmission of a status update requires multiple channel uses, the authors propose a transmission strategy to decide if an ongoing transmission should be aborted when a new status update arrives. In [16], the authors find sampling strategies for a Wiener process, optimizing the AoI by using a threshold strategy.

Compared to the extensive work on AoI, there have been only initial contributions on other age-based metrics. After motivating and introducing the AoII, the authors of [17] optimize the average AoII. First, a point-to-point scenario with unlimited energy usage is examined, followed by an exploration of a more realistic setting with energy constraints. A similar energy-constrained scenario for the AoII is considered in [18], where the authors use Relative Value Iteration to find an AoII-optimal policy. In [4], the average QAoI is introduced and optimized for the point-to-point scenario and different types of queries by using policy iteration. In [19] and [20], the QAoI is optimized by implementing a pull-based strategy. In this case, the receiver node decides the time of an update.

The aforementioned works focus on minimizing age-based metrics on average, i.e., they optimize these metrics over a long-term time horizon. However, minimizing the average AoI, AoII, or QAoI does not ensure the prevention of cases where the age-based metric experiences exceptionally high values. For this reason, a new research direction has emerged which, in addition to minimizing the average AoI, focuses on peak AoI. Most current works considering peak AoI characterize the probability of reaching high AoI values under different assumptions, e.g., short status update packets [12], status update sources with and without retransmissions [13], and customizable status update arrival rates at the sender [14]. As previously explained, many applications require risk-sensitivity. However, the design of risk-aware transmission strategies at the sender has, so far, received little attention. This holds for the AoI and in particular for its variants, the AoII and the QAoI. The authors in [15] take a step in this direction by using value iteration to derive a risk-aware transmission strategy at the sender for the AoI when the probabilities for a status update arrival and for a successful transmission are apriori known. Although value iteration leads to the optimal transmission policy, it lacks scalability. It is computationally expensive and can only be applied to derive the optimal policy in a reasonable time for small sets of possible values of the AoI. In [21], the authors discuss the trade-off between energy and AoI, while also analyzing peak AoI. They propose a Truncated Automatic Repeat reQuest (TARQ) strategy, which sends new status updates until they are successfully transmitted, but only until a predefined number of attempts is reached. However, this strategy is suboptimal as it does not take the AoI at the receiver into account. The authors of [22] explore non-linear age metrics derived from the AoI and information-based data freshness metrics similar to the AoII. They address the sampling problem in which the sender generates updates at its discretion and transmits them directly over a channel modeled as a FIFO-queue, distinguishing their problem from the one considered here.

| Age-Based Metric | Risk-Neutral | Risk-Sensitive |
|---|---|---|
| Age of Information | [8]–[11] | [12]–[15], [21] |
| Query Age of Information | [4], [19], [20] | - |
| Age of Incorrect Information | [5], [18] | - |

TABLE I: Related work on optimizing age-based metrics.

Our objectives encompass three fundamental aspects: reducing the values of age-based metrics on average, minimizing the required transmission energy on average, and mitigating the risk of experiencing high age-based metric values. To address these objectives, we introduce the concept of a *risky state*, i.e., the case in which the considered age-based metric is higher than a predefined safety value. From this definition, we derive the *frequency of risky states* as a risk-metric and leverage this notion to *quantify* and *minimize* the risk of encountering large values of age-based metrics. The advantages of this new concept lie in its simplicity, its general applicability to all considered metrics, and its potential to enhance the development of new risk-sensitive transmission strategies for critical applications. In contrast to the existing works, we consider all the age-based metrics discussed above, namely AoI, QAoI, and AoII, in order to design risk-sensitive transmission strategies for point-to-point wireless communication scenarios. Table I gives an overview and differentiates related works on age-based metrics into risk-neutral and risk-sensitive categories. With our focus on risk within these metrics, we aim to fill the evident research gap. We use our notion of risky states to propose two different approaches to develop risk-sensitive transmission strategies which are applicable to all the considered age-based metrics and also in applications with large sets of possible values of the age-based metrics. The first approach is dependent on knowledge of system parameters, e.g., the channel quality, which allows us to use analytical methods to find the optimal strategy. The second approach overcomes this requirement by relying on learning.

In considering energy and age, our approaches assess an average cost per time step. The cost is defined as a weighted sum of the specific age-based metric at the receiver and the transmission energy used at the sender. To strike a balance between the cost and the frequency of risky states, our strategies incorporate parameters that we call *risk parameters*. To assess our strategies, we separately analyze both risk and cost metrics instead of using aggregated metrics like weighted sums. This methodology not only allows us to better understand the effectiveness of each strategy but also to set them against other baseline methods. Through this detailed evaluation, we extract valuable insights into how well these strategies manage risk and cost.

Our contributions can be summarized as follows:

- We define *risky states* for which an age-based metric exceeds a safety threshold. Using this, we formulate the *frequency of risky states* as a risk metric to quantify and minimize associated risks.
- For AoI and QAoI, we propose a threshold-based trans-

mission strategy and introduce a tunable parameter called the *transmission threshold*. The transmission of status updates is triggered whenever the difference between the AoI (or QAoI) at the sender and at the receiver exceeds this threshold. By tuning this parameter, the strategy can meet various demands for different applications. For instance, lower thresholds lead to reduced risk, but may also result in higher costs. We provide a mathematical derivation of the costwise optimal threshold and derive a closed-form expression for the strategy's average cost. We further provide a closed-form expression for the frequency with which risky states are visited under a threshold-based strategy. This expression is used to choose a transmission threshold guaranteeing a low frequency of risky states.

- For the AoII, the derivation of the closed-form expression for the strategy's average cost used for AoI and QAoI is not applicable. We hence find the optimal threshold-based strategies empirically. In contrast to AoI and QAoI, measuring the AoII assumes current knowledge about the state of the underlying process at the sender at all times, such that the threshold is applied to the AoII itself instead of the difference between the value at the sender and at the receiver. We show that this empirical search for the optimal threshold can be carried out in a risk-aware manner by taking into account an upper bound for the probability of risky states.
- When calculating the optimal threshold for the threshold-based strategy, knowledge of the relevant probabilities used in the model, e.g., status update arrival probability or the probability of a successful transmission is needed. To drop these requirements, we propose a novel risk-sensitive variation of $Q$-learning. We directly include our proposed notion of risky states in a risk-aware learning algorithm, termed *$Q$-learning using risky states* ($Q$+RS). $Q$+RS is able to balance cost and risk using a tunable risk-parameter for all of the considered age-based metrics. At the same time, $Q$+RS does not depend on apriori knowledge of the probabilities of a new status update arrival and of a successful transmission. By means of numerical simulations, we show that, compared to traditional $Q$-learning, $Q$+RS can not only reduce the occurrence of risky states but also the cost in the system.

Briefly summarizing the extensions compared to our conference papers [23] and [24], here, we have significantly extend our analysis. The key extensions compared to these papers are as follows: For the first time, we address both the QAoI and the AoII in the context of risk, establishing a comprehensive framework for evaluating risk across these metrics. We provide an extended version of the proof detailing the cost implications of the threshold-based strategy for the AoI, which is then adapted for the QAoI. Additionally, we extend the proof concerning the frequency of risky states under the threshold-based strategy for the AoI, and similarly transfer this proof to the QAoI. Utilizing these proofs, we demonstrate that our method, initially proposed for the AoI, is equally applicable to the QAoI. We significantly expand the numerical simulations to include an extended analysis of the second method proposed in [24], which is based on $Q$-learning, to examine its behavior for a higher number of training time steps. For the QAoI, we not only provide a learning algorithm to find the optimal threshold but also a proof demonstrating how to determine the optimal threshold without simulations. In comparison to [23], where we performed a risk-neutral empirical threshold search, we now also explain how to render this search risk-aware for both QAoI and AoII. We extend the simulations for AoII and QAoI to the same scenario used in [24] and include simulations to visualize the search for the optimal threshold for AoI, QAoI, and AoII. Furthermore, we include simulations to demonstrate the impact of different query probabilities for the QAoI. Additionally, we have incorporated a new comparison strategy *TARQ* in the extended simulations.

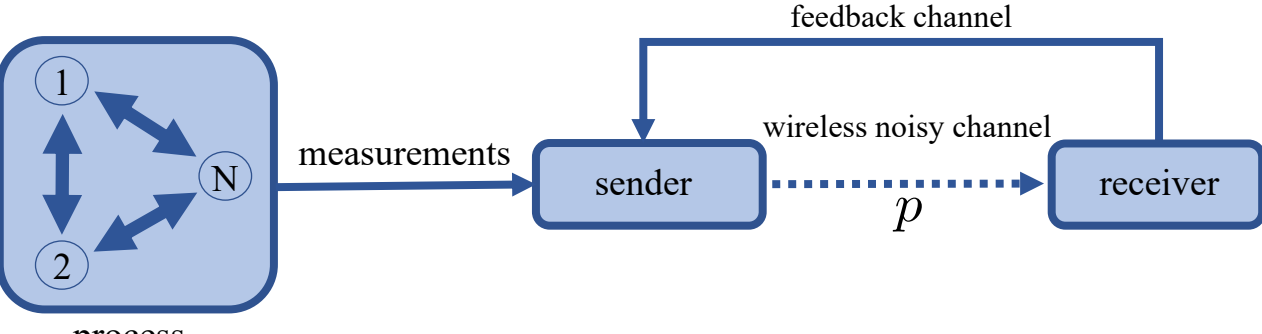

Fig. 1: System Model

The rest of the paper is organized as follows: In Section II, we present the considered system model. The optimization problem along with our threshold-based solution for AoI is introduced in Section III. In Sections IV and V, we present the respective optimization problem and the threshold-based solutions, first for QAoI and then for AoII, respectively. Our new risk-aware learning solution, $Q$+RS, is introduced in Section VI. In Section VII, we provide the numerical evaluation of the proposed strategies. Finally, Section VIII concludes the paper.

## II. System Model

Similar to [15] and as depicted in Fig. 1, the analyzed system comprises an underlying process observed by an IoT device, denoted as the sender, along with a receiver and a wireless packet erasure channel connecting them. The IoT device receives status updates from the underlying process. The receiver relies on up-to-date information from this IoT device, utilizing it as input for data-driven tasks.

As in [5], we model the underlying process as a Markov chain with $N$ states. The probability of remaining in a state is the same for all states and is called $p_r$. The probability of changing to a certain other state is again the same for each transition and is denoted by $p_c$. The resulting relationship is

$$p_r + (N-1)p_c = 1. \tag{1}$$

The system uses discrete and equidistant time steps, indexed by $t \in \mathbb{N}$. The status update arrival process at the sender is modeled as a Bernoulli process, such that at the beginning of each time step $t$, an update arrives at the sender with probability $\lambda$. The sender has a buffer, able to store only the freshest status update. This means that, as soon as a new status update arrives, the currently stored update is replaced by the new one. In each time step $t$, the sender has then to decide whether it wants to send the currently stored status update to the receiver or not. If the sender decides to transmit, the status update can correctly be detected at the receiver with a probability of $p$. This probability $p$ models the quality of the wireless noisy channel. We assume that each sending attempt uses the same amount of energy $\nu$. We further assume that the sender receives information on whether the packet is correctly detected at the receiver or not via a perfect feedback channel.

Let $\mathrm{A}_{\mathrm{Rx},t}$ denote any of the age-based metrics, i.e., AoI, QAoI and AoII, at the receiver at time $t$. The sender's decision to transmit or not in time step $t$ results in a cost $C_t$ *associated with the single time step* $t$. $C_t$ is defined as the weighted sum of the considered age-based metric $\mathrm{A}_{\mathrm{Rx},t}$ at the receiver, and the transmission energy $\nu$. Formally, $C_t$ is defined as

$$C_t = \begin{cases} \alpha \mathrm{A}_{\mathrm{Rx},t} + \beta\nu & \text{if the sender sends,} \\ \alpha \mathrm{A}_{\mathrm{Rx},t} & \text{otherwise,} \end{cases} \tag{2}$$

where $\alpha$ and $\beta$ are weights on the age-based metric and the energy cost. The cost *of a strategy* $\pi$ is defined to be the long-term average of the costs in all single time steps:

$$\mathit{cost}(\pi) := \lim_{T\to\infty} \frac{1}{T} \sum_{t=1}^{T} \mathbb{E}[C_t|\pi], \tag{3}$$

where $\mathbb{E}[C_t|\pi]$ denotes the expected costs in time step $t$ under strategy $\pi$. The expected value is computed within the probability space that models the instances of successful transmissions, the arrival of new packets, and the posing of queries.

We additionally introduce the concept of *risky states* as events in which the age-based metric exceeds a predefined safety value $\zeta \in \mathbb{N}$, i.e. in which $\mathrm{A}_{\mathrm{Rx},t} \geq \zeta$. In the context of the given application, this safety value $\zeta$ quantifies the notion that the information at the receiver is significantly outdated, leading to considerable uncertainty about the latest data on the observed process, which can result in high risk. For example, this could lead to the inability to maintain control cycles, posing potential disruptions and challenges to system stability.

## III. Age of Information

### A. AoI Definition

In this section, we first define the AoI. We then state the problem formulation for the balancing problem between AoI and transmission energy. Finally, we present our threshold-based solution for this problem.

Assuming the status updates arrive at random time steps $t = t_i$, where $i \in \mathbb{N}$, the AoI at the sender evaluated at time step $t \in [t_i, t_{i+1} - 1]$, denoted as $\mathrm{AoI}_{\mathrm{Tx},t} \in \mathbb{N}_0$, is defined as

$$\mathrm{AoI}_{\mathrm{Tx},t} := t - t_i, \quad \text{for } t \in [t_i, t_{i+1} - 1]. \tag{4}$$

We additionally set $t_1 := 1$, resulting in $\mathrm{AoI}_{\mathrm{Tx},1} = 0$. Hence, the minimal $\mathrm{AoI}_{\mathrm{Tx},t}$ is 0. An example of a possible timeline of $\mathrm{AoI}_{\mathrm{Tx},t}$ is shown in Fig. 2. Here, updates arrive at the beginning of the time steps indexed with $1, 5, 6$, and $11$, such that $t_1 = 1$, $t_2 = 5$, $t_3 = 6$ and $t_4 = 11$. At these time steps, $\mathrm{AoI}_{\mathrm{Tx},t}$ is set to 0. In all the remaining time steps, $\mathrm{AoI}_{\mathrm{Tx},t}$ is increased by 1.

The AoI at the receiver, denoted as $\mathrm{AoI}_{\mathrm{Rx},t} \in \mathbb{N}$, is defined as

$$\mathrm{AoI}_{\mathrm{Rx},t+1} := \begin{cases} \mathrm{AoI}_{\mathrm{Tx},t} + 1, & \text{if a transmission} \\ & \text{attempt succeeds at } t, \\ \mathrm{AoI}_{\mathrm{Rx},t} + 1, & \text{otherwise.} \end{cases} \tag{5}$$

Note that as in [15], the lowest possible value of $\mathrm{AoI}_{\mathrm{Rx}}$ is 1, while the lowest possible value of $\mathrm{AoI}_{\mathrm{Tx}}$ is 0. Moreover, we set $\mathrm{AoI}_{\mathrm{Rx},1} := 1$.

### B. Problem Formulation

Our goal is to design a transmission strategy $\pi$ at the sender that minimizes the cost defined in (3). This problem can be formulated as an average-cost Markov Decision Process (MDP) $\mathcal{M}$. For given parameters $p, \lambda \in (0,1)$ and $\nu \geq 0$, the MDP $\mathcal{M}$ modeling the described system consists of a set $\mathcal{S} := \mathbb{N}_0 \times \mathbb{N}$ of states, a set $\mathcal{A} := \{0,1\}$ of actions, a cost function $c$ and state transition probabilities given by a function $P$. Each state $s \in \mathcal{S}$ is a pair of natural numbers modeling the AoI at the sender and at the receiver, i.e., $s = (\mathrm{AoI}_{\mathrm{Tx}}, \mathrm{AoI}_{\mathrm{Rx}})$. The action space $\mathcal{A}$ contains two actions. Action 0 means that the sender waits and does not transmit the status update from its buffer. Action 1 corresponds to a sending attempt. The cost function $c$ returns the cost of a state-transition $(s_t, a, s_{t+1})$, i.e., the cost arising from transitioning from state $s_t = (\mathrm{AoI}_{\mathrm{Tx},t}, \mathrm{AoI}_{\mathrm{Rx},t})$, at time step $t$, to state $s_{t+1} = (\mathrm{AoI}_{\mathrm{Tx},t+1}, \mathrm{AoI}_{\mathrm{Rx},t+1}) \in \mathcal{S}$ at time step $t+1$ after taking action $a \in \mathcal{A}$. We define the function $c : \mathcal{S} \times \mathcal{A} \times \mathcal{S} \to \mathbb{R}$ as $c(s_t, a, s_{t+1}) = C_{t+1}$ using $C_{t+1}$ defined in (2). According to the previously described system, the transition probability function $P : \mathcal{S} \times \mathcal{A} \times \mathcal{S} \to [0,1]$ is defined as

$$P(s_t, 0, s_{t+1}) := \tag{6}$$
$$\begin{cases} \lambda, & \text{if } s_{t+1} = (0, \mathrm{AoI}_{\mathrm{Rx},t} + 1) \\ 1 - \lambda, & \text{if } s_{t+1} = (\mathrm{AoI}_{\mathrm{Tx},t} + 1, \mathrm{AoI}_{\mathrm{Rx},t} + 1), \end{cases}$$

$$P(s_t, 1, s_{t+1}) := \tag{7}$$
$$\begin{cases} p\lambda, & \text{if } s_{t+1} = (0, \mathrm{AoI}_{\mathrm{Tx},t} + 1) \\ p(1-\lambda), & \text{if } s_{t+1} = (\mathrm{AoI}_{\mathrm{Tx},t} + 1, \mathrm{AoI}_{\mathrm{Tx},t} + 1) \\ (1-p)\lambda, & \text{if } s_{t+1} = (0, \mathrm{AoI}_{\mathrm{Rx},t} + 1) \\ (1-p)(1-\lambda), & \text{if } s_{t+1} = (\mathrm{AoI}_{\mathrm{Tx},t} + 1, \mathrm{AoI}_{\mathrm{Rx},t} + 1). \end{cases}$$

Strategies $\pi$ for the solution of this MDP are maps from $\mathcal{S}$ to $\mathcal{A}$. Expressing the average cost as in (3) for the MDP, we write

$$\mathit{cost}(\pi) := \lim_{T\to\infty} \frac{1}{T} \sum_{t=1}^{T} \mathbb{E}[c(s_t, \pi(s_t), s_{t+1})], \tag{8}$$

where the occurrence of the state $s_t$ at time step $t$ depends on the transition probabilities of $\mathcal{M}$. $\mathbb{E}[c(s_t, \pi(s_t), s_{t+1})]$ is the expected cost of the transition $(s_t, \pi(s_t), s_{t+1})$. Additionally, for a safety value $\zeta$, the set $\mathcal{R}$ of risky states is given as

$$\mathcal{R} := \{s = (\mathrm{AoI}_{\mathrm{Tx}}, \mathrm{AoI}_{\mathrm{Rx}}) \in \mathcal{S} | \mathrm{AoI}_{\mathrm{Rx}} \geq \zeta\}. \tag{9}$$

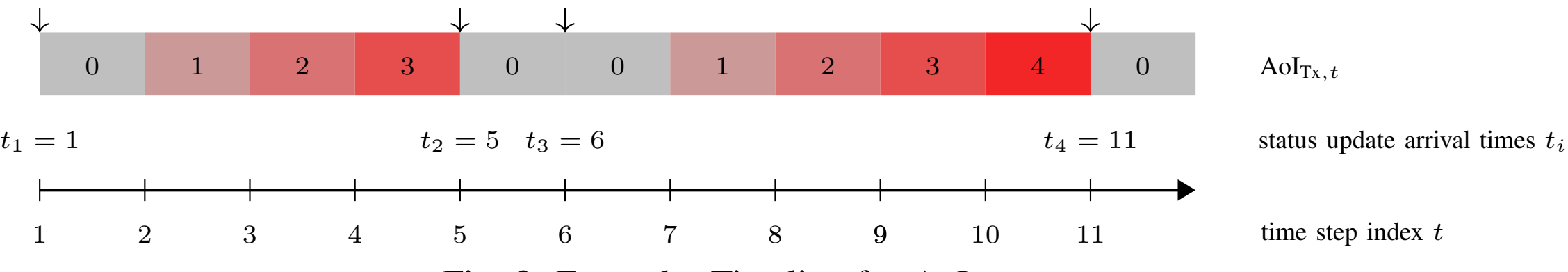

Fig. 2: Example: Timeline for $\mathrm{AoI}_{\mathrm{Tx},t}$

Note that in this problem formulation, we do not assume specific dynamics of the underlying process. This allows to drop the assumption that the process has to be modeled as a Markov chain for the case of AoI. While this is not possible for the AoII, here, it allows for a more general view of the process, which, e.g. does not need to fulfill the Markov property.

### C. Threshold-Based Approach

Here, we introduce our proposed threshold-based transmission strategy, which is developed through optimization. The core insight behind this strategy is that a larger difference between $\mathrm{AoI}_{\mathrm{Rx}}$ and $\mathrm{AoI}_{\mathrm{Tx}}$ indicates a greater potential advantage from attempting a transmission.

We start by considering the risk-neutral scenario as described in the above problem formulation. Subsequently, we outline the process of adapting the threshold-based approach to incorporate risk-sensitivity. First, we determine the costwise optimal threshold $n$ for the difference between $\mathrm{AoI}_{\mathrm{Rx},t}$ and $\mathrm{AoI}_{\mathrm{Tx},t}$. This threshold serves as a decision boundary, where the transmitter will wait if the difference is below $n$ and send data when the difference is equal to or exceeds $n$.

The threshold-based strategy $\pi_{TB}(n)$ is hence characterized by a threshold $n$. According to $\pi_{TB}(n)$, the transmitter sends, if and only if the difference between the AoI at the receiver and that at the sender is equal to or larger than $n$. We define $\pi_{TB}(n)$ as the following map from $\mathcal{S}$ to $\mathcal{A}$:

$$\pi_{TB}(n)((\mathrm{AoI}_{\mathrm{Tx}}, \mathrm{AoI}_{\mathrm{Rx}})) = \begin{cases} 0 & \text{for } \mathrm{AoI}_{\mathrm{Rx}} - \mathrm{AoI}_{\mathrm{Tx}} < n, \\ 1 & \text{for } \mathrm{AoI}_{\mathrm{Rx}} - \mathrm{AoI}_{\mathrm{Tx}} \geq n. \end{cases} \tag{10}$$

The intuitive idea behind the threshold-based strategy is that in case of a successful transmission, the $\mathrm{AoI}_{\mathrm{Rx}}$ is reduced by the difference $\mathrm{AoI}_{\mathrm{Rx},t} - \mathrm{AoI}_{\mathrm{Tx},t}$, i.e., a decision to send is more profitable for a higher difference.

We continue with a lemma about the cost $cost(\pi_{TB}(n))$ associated with the strategy $\pi_{TB}(n)$. This lemma is used to find the costwise optimal value for the threshold $n$ and to derive the costwise optimal threshold-based strategy TB-baseline. Afterwards, risk is considered in Lemma 3, where we provide a term for the frequency of the appearance of risky states during the strategy's execution. Combining both lemmas, we are able to find a value for the threshold $n$ optimizing the cost under a given risk constraint. To apply the lemmas, the risk constraint has to be given in terms of a maximal frequency for the appearance of risky states.

**Lemma 1.** *The average cost of the strategy $\pi_{TB}(n)$ is*

$$cost(\pi_{TB}(n)) = \frac{\alpha\lambda p \left( \sum_{r=1}^{n} w_r(a(n) - \frac{r(r-1)}{2}) + \sum_{r=n+1}^{\infty} w_r(r + a(1)) \right) + \beta\lambda\nu}{\lambda(1-p) + 1 + p(1-\lambda) + \sum_{r=1}^{n-1} w_r(n-r)}, \tag{11}$$

*where*

$$a(n) := \frac{n-2}{\lambda} + \frac{n-2}{p} + \frac{(n-2)(n-1)}{2} + \frac{1}{\lambda^2} + \frac{1}{\lambda p} + \frac{1}{p^2} \tag{12}$$

*and*

$$w_r := (1-\lambda)^{r-1}(1-p)^{r-1} - (1-\lambda)^r(1-p)^r. \tag{13}$$

*Proof.* The intuition behind the proof is to first define a *period* as the time between two successful transmission attempts. By calculating the average cost of each period and combining the individual results, we can prove the above formula for $cost(\pi_{TB}(n))$.

The detailed proof can be found in Appendix A. □

To find the threshold for the costwise optimal threshold-based strategy TB-baseline, the resulting term for $cost(\pi_{TB}(n))$ from Lemma 1 can be easily minimized in $n$. This is because the corresponding function in $n$ is convex in the considered parameter space. A corresponding proof is sketched in Appendix C.

To illustrate the influence of parameter variations on the optimal threshold $n$, we present a consideration of the boundary cases. Holding $\lambda$, $p$, and $\nu$ constant, the dynamics between $\alpha$ and $\beta$ dictate the behaviour of the cost. As $\alpha$ approaches infinity and $\beta$ reduces to zero, it is favorable to frequently transmit due to the lower relative cost of transmissions compared to the increased AoI, thus rendering $n = 1$ as the optimal threshold. In this case $cost(\pi_{TB}(n))$ will increase in $n$ for $n \geq 1$. Conversely, with $\alpha$ approaching zero and $\beta$ increasing towards infinity, the high transmission costs compared to AoI costs result in a larger optimal threshold for $n$. Here, $cost(\pi_{TB}(n))$ will decrease in $n$ until this large optimal threshold is reached.

Please note that TB-baseline is originally designed in a risk-neutral manner. However, we can incorporate risk into the threshold-based strategy by using a lower threshold than the costwise optimal threshold. The key challenge remains in calibrating this threshold to strike a balance between risk sensitivity and cost efficiency.

We provide an expression to determine the frequency of *risky states* with high $\mathrm{AoI}_{\mathrm{Rx}}$ during the execution of the strategy. By evaluating this expression, we can find a sufficiently low threshold to meet a risk constraint, which limits the maximal frequency of risky states. Note that this is only possible if the risk constraint can be satisfied by any strategy. On the other hand, if a decision threshold is given, this expression enables us to quantify the associated risk. To make the notion of a frequency precise, we use the following definition.

**Definition 2.** *For a sequence of random variables $(s_i)_{i=1,2,\ldots}$, the frequency $f_A$ of an event $A$ is defined as*

$$f_A := \mathbb{E}[\lim_{T\to\infty} \frac{1}{T} \sum_{i=1}^{T} \mathbb{1}_{s_i \in A}]. \tag{14}$$

For the threshold-based strategy $\pi_{TB}(n)$, risky states with $\mathrm{AoI}_{\mathrm{Rx}} = k \geq \zeta$ appear with the frequency $f_k$ given by the following lemma. Note that the lemma holds for all $\mathrm{AoI}_{\mathrm{Rx}} = k > n$, where $n$ is the strategy's transmission threshold.

**Lemma 3.** *For the strategy $\pi_{TB}(n)$, the frequency $f_k$ of an AoI at the receiver of $k > n$ is given by*

$$f_k = \frac{w_k + P_n \cdot (\sum_{r=1}^{n} w_r) + \sum_{r=n+1}^{k-1} w_r P_r}{l}, \tag{15}$$

*where*

$$w_r = (1-\lambda)^{r-1}(1-p)^{r-1} - (1-\lambda)^r(1-p)^r \tag{16}$$

*and*

$$P_r := 1 - p\lambda \sum_{j=0}^{k-r-1} \sum_{j=0}^{i} (1-\lambda)^j (1-p)^{i-j} \tag{17}$$

*and*

$$l = \frac{1-p}{p} + 1 + \frac{1-\lambda}{\lambda} + \sum_{r=1}^{n-1} w_r \cdot (n-r). \tag{18}$$

*Proof.* This proof follows a similar approach to the proof of Lemma 1. Again, we use periods defined as the time between two successful transmission attempts and calculate the individual frequency of risky states for each period. Combining the individual results yields the above formula.

The detailed proof can be found in Appendix B. □

## IV. Query Age of Information

### A. QAoI Definition

In this section, we start with the definition of the QAoI, state the corresponding problem formulation, and discuss the applicability of the threshold-based approach described in Sec. III-C. QAoI is useful in cases in which the receiver is only interested in specific query time steps. To consider QAoI, the cost defined in (3) needs to be modified. Specifically, we define the costs for QAoI as in [4]:

$$\begin{aligned} cost_{QAoI}(\pi) := & \lim_{T\to\infty} \frac{1}{T} \sum_{\substack{t\in Q \\ t\leq T}} \mathbb{E}[\alpha \mathrm{AoI}_{\mathrm{Rx},t} | \pi] \\ & + \lim_{T\to\infty} \sum_{t\leq T} \mathbb{E}[\beta \nu a_t | \pi], \end{aligned} \tag{19}$$

where $Q \subseteq \mathbb{N}$ is the set of *query time steps*. Examples for $Q$ are periodic time steps with period $k \in \mathbb{N}$ ($Q = \{kt, t \in \mathbb{N}\}$) or stochastic queries, where each time step has a probability $q \in [0,1]$ to be a query time step ($\mathbb{P}(t \in Q) = q$ for each $t \in \mathbb{N}$).

### B. Problem Formulation

Using (19), the problem can be formulated using the MDP $\mathcal{M}$ described for AoI in Sec. III-B. Also, risky states are defined accordingly as

$$\mathcal{R} := \{s_t = (\mathrm{AoI}_{\mathrm{Tx},t}, \mathrm{AoI}_{\mathrm{Rx},t}) \in \mathcal{S} | \mathrm{AoI}_{\mathrm{Rx}} \geq \zeta \wedge t \in Q\}. \tag{20}$$

Note that as for AoI, we do not assume specific dynamics of the underlying process. This again allows us to drop the assumption that the process has to be modeled as a Markov chain for the case of QAoI.

### C. Threshold-Based Approach

By employing the assumption stated in [4] that each time step is considered a query time step independently of all other time steps and regardless of the outcome of other relevant random variables with a probability of $q$, we can readily derive an explicit expression for the cost of a threshold-based strategy $\pi_{TB}(n)$ concerning QAoI. We then get

**Lemma 4.** *Assume that $Q \subseteq \mathbb{N}$ contains every natural number independently with a probability of $q$. Measuring the QAoI for this set of query time steps $Q$, the average cost of the strategy $\pi_{TB}(n)$ is given by*

$$cost_{QAoI}(\pi_{TB}(n)) = \frac{q\alpha\lambda p \left( \sum_{r=1}^{n} w_r \left( a(n) - \frac{r(r-1)}{2} \right) + \sum_{r=n+1}^{\infty} w_r (r + a(1)) \right) + \beta\lambda\nu}{\lambda(1-p) + 1 + p(1-\lambda) + \sum_{r=1}^{n-1} w_r (n-r)}, \tag{21}$$

*where $a(n)$ and $w_r$ are defined as in Lemma 1.*

*Proof.* The proof is similar to the proof for the AoI case in Lemma 1. A detailed proof can be found in Appendix D. □

Please note that Lemma 3 can be seamlessly applied to QAoI by simply multiplying the resulting frequencies with $q$. This allows for a consistent treatment of risk, mirroring our proposed approach for AoI.

## V. Age of Incorrect Information

### A. AoII Definition

Following the same structure as in Sec. IV, in this section, we define AoII, state the corresponding problem formulation and discuss the applicability of the threshold-based approach.

In contrast to AoI and QAoI, the AoII takes the content of sent packets into account. It is formally defined as:

$$\mathrm{AoII}_{t+1} := \begin{cases} 0 & \text{if the information at the receiver} \\ & \text{about the process is correct} \\ \mathrm{AoII}_t + 1, & \text{otherwise.} \end{cases} \tag{22}$$

We additionally set $\mathrm{AoII}_1 := 0$.

### B. Problem Formulation

As the packet content is relevant for AoII, in this section we adjust the MDP $\mathcal{M}$ introduced in Sec. III-B. We call the new process $\mathcal{M}_{AoII} := (\mathcal{S}_{AoII}, \mathcal{A}, c_{AoII}, P_{AoII})$. While the set of actions $\mathcal{A} = \{0, 1\}$ remains the same, the set of states has to be adjusted. Recall that it is necessary for the sender to have current knowledge about the state to calculate the AoII. Therefore, a status update has to arrive at the sender at every time step (and the probability of the Bernoulli arrival model has to be fixed at $p = 1$).

As the AoI at the transmitter is now constantly 0, we drop its value from the state space. Instead, we add a variable, which is 0 if the state of the underlying process does not equal the information at the receiver, and 1 if the information matches the current process state. The AoI at the receiver will be exchanged by the AoII. This results in $\mathcal{S} := \{0, 1\} \times \mathbb{N}_0$, where the single states $s \in \mathcal{S}$ are either $s = (0, AoII)$ or $s = (1, AoII)$. We define the new cost function as

$$cost_{AoII}(\pi) := \lim_{T\to\infty} \frac{1}{T} \sum_{t=1}^{T} \mathbb{E}[C_t^{AoII} | \pi], \tag{23}$$

where

$$C_t^{AoII} = \begin{cases} \alpha \text{AoII}_t + \beta\nu & \text{if the sender sends,} \\ \alpha \text{AoII}_t & \text{otherwise.} \end{cases} \tag{24}$$

It is important to note that in contrast to the $\text{AoI}_{\text{Rx}}$, the AoII checks the correctness of the information at the receiver *after* a possible sending attempt. In case of a successful attempt it is therefore set to 0 instead of 1. The $\text{AoI}_{\text{Rx}}$ as defined in Equation (5) checks whether the sending attempt in the *previous* time step was successful and will therefore remain greater than 0 in every time step. For $\mathcal{M}_{AoII}$ it remains to define $c_{AoII}(s_t, a, s_{t+1}) := C_{t+1}^{AoII}$ and the transition probabilities $P_{AoII} : \mathcal{S}_{AoII} \times \mathcal{A} \times \mathcal{S}_{AoII} \to [0, 1]$:

$$P_{AoII}(s_t, 0, s_{t+1}) := \begin{cases} p_r & s_t = (1, 0) \wedge s_{t+1} = (1, 0), \\ 1 - p_r & s_t = (1, 0) \wedge s_{t+1} = (0, 1), \\ p_c & s_t = (0, x) \wedge s_{t+1} = (1, 0), \\ 1 - p_c & s_t = (0, x) \wedge s_{t+1} = (0, x + 1), \end{cases}$$

$$P_{AoII}(s_t, 1, s_{t+1}) := \begin{cases} p_r + p \cdot (1 - p_r) & s_t = (1, 0) \wedge s_{t+1} = (1, 0), \\ 1 - (p_r + p \cdot (1 - p_r)) & s_t = (1, 0) \wedge s_{t+1} = (0, 1), \\ p_c + p \cdot (1 - p_c) & s_t = (0, x) \wedge s_{t+1} = (1, 0), \\ 1 - (p_c + p \cdot (1 - p_c)) & s_t = (0, x) \wedge s_{t+1} = (0, x + 1), \end{cases} \tag{25}$$

where $\wedge$ is the logical conjunction operation and $x \in \mathbb{N}$. Similar to risky states for the AoI, we define the set of risky states for the AoII as

$$\mathcal{R}_{AoII} := \{s \in \mathcal{S} \text{ with AoII}(s) \geq \zeta_{AoII}\}, \tag{26}$$

where $\zeta_{AoII}$ is a given safety value and $\text{AoII}(s)$ is the second entry of the state $s$. Note that $\zeta_{AoII}$ can differ from $\zeta$ in the original MDP designed for the AoI, even for the same underlying scenario. This is because the application's restrictions on the AoI might be different from the application's restrictions on the AoII.

### C. Threshold-Based Approach

As the AoII is based on $\mathcal{M}_{AoII}$ instead of $\mathcal{M}$, it is not possible to directly transfer the proof of Lemma 1. Picking up the idea of threshold-based solutions, we present here an empiric approach to finding the best transmit threshold for the AoII. This approach does not necessarily find an optimal solution, but we can use it as a baseline and for comparison with our learning algorithms of Sec. VI.

For the AoI, we found the optimal threshold of the difference of the AoI at the receiver and at the sender. For the AoII, we find a threshold not for this difference but simply for the AoII-value itself. To find this threshold, we run a simulation of $\mathcal{M}_{AoII}$ for a predefined set of threshold-based strategies several times. We then pick the threshold-based strategy with the smallest average cost. This empiric search can also be used to illustrate the result of Lemma 1 and Lemma 4, as we can find that the optimal threshold indeed generates the smallest average cost out of all threshold-based strategies. If it is not only necessary to optimize the cost but also to visit a small number of risky states, the approach can be adjusted. As the frequency of risky states is increasing with higher thresholds, the search can be stopped as soon as a threshold-based strategy is tested which visits risky states with a higher frequency than permitted. Afterward, we can pick the strategy with the lowest average costs which does not exceed the given bound for the frequency of risky states.

## VI. $Q$-Learning Based Approach

In this section, we present the risk-sensitive learning algorithm $Q$+RS, which combines $Q$-learning and the notion of *risky states*. $Q$+RS does not depend on any knowledge of the system parameters. $Q$+RS is also not limited to small sets of possible AoIs as the value iteration approach in [15], because in contrast to value iteration, the number of performed machine operations does not grow in the size of the state space.

We apply $\epsilon$-greedy tabular $Q$-learning to the MDP in Sec. III-B. $Q$-learning in its original form is risk-neutral in the sense that it optimizes costs in the MDP without considering any risk-metric. In contrast to this original form of $Q$-learning, we achieve risk-sensitivity by modifying the cost function $c$ given by the MDP.

The use of a modified cost function $c$ can be naturally combined with the notion of *risky states*. This is achieved by multiplying costs for transitions to *risky states* by a risk factor $\rho > 1$. If $\rho > 1$ the algorithm learns risk-sensitive strategies. $\rho = 1$ results in the original MDP, while $\rho < 1$ would result in risk-seeking strategies.

We opt to include risk in this way because it is remarkably simple, perfectly transferable between AoI, QAoI, and AoII, inherits all convergence properties of $Q$-learning, and demonstrates strong performance in practice (see Sec. VII). Also, in this way, risk is considered at every time step, not only when a risky state might appear immediately. This is a natural property of learning algorithms like $Q$-learning, as they "look into the future" and adjust their strategy based on potential future states.

Note that identifying risky states is an important part of this approach. Therefore, we require the application to provide us with guidance on how old the information should be at the maximum (in other words, we need the application to provide a $\zeta$, such that we can derive the set $\mathcal{R}$).

The modified cost function as implemented in Algorithm 1 is then defined as

$$c_{\mathcal{R}}(s, a, s') := (\mathbb{1}_{s' \notin \mathcal{R}} + \rho \cdot \mathbb{1}_{s' \in \mathcal{R}}) \cdot c(s, a, s'). \tag{27}$$

Note that this modified cost function is the difference between traditional $Q$-learning and $Q$+RS. The original cost function of TQL can be regained by setting $\rho = 1$.

The pseudo-code for $Q$+RS is given in Algorithm 1. The algorithm iteratively approximates the so-called $Q$-value of each state-action pair, i.e., the pair's expected future cost. The resulting approximations after $N$ iterations are called $Q^{(N)}$-values. The initial approximations $Q^{(0)}$ are set to be 0. After initial operations (lines 1-2), the algorithm works in an iterative fashion (l. 3-18). The $\epsilon$-greedy strategy used during learning chooses a random action with a probability of $\epsilon$ and the action with the lowest estimated $Q$-value with a probability of $(1-\epsilon)$ (l. 4). During learning, $\epsilon$ is reduced by multiplying it by a decay factor $\delta \in (0, 1)$ after every iteration (l. 5). The $Q$-value update from traditional $Q$-learning (l. 10-16) is used with an additional manipulation of the time step's cost $C_{t+1}$ (l. 7-9) in case of risky states. To weigh current and future costs (see l. 12), $Q$-learning uses a discount factor $\gamma$, which we here introduce as a hyperparameter, a predefined configuration variable that influences the learning process but is not learned from the data. From the resulting $Q^{(N)}$-values, a strategy is constructed by choosing the action with the lowest $Q^{(N)}$-value in each state.

This algorithm can easily be adjusted to variations of the AoI. For the QAoI, we exchange $C_{t+1}$ by 0 for non-query time steps. To find $Q$-values for the AoII, we exchange the MDP $\mathcal{M}$ by the MDP $\mathcal{M}_{AoII}$, the starting state by $s_{0,AoII} = (0, 0)$ and the set of risky states by $\mathcal{R}_{AoII}$.

Note that in contrast to the threshold-based approaches above, for Q-learning, we do not provide a guarantee about the probability of encountering risky states.

**Algorithm 1:** $Q$-learning + risky states ($Q$+RS)

**Data:** simulator for $\mathcal{M}$, starting state $s_0$, no. of time steps $N$, actions $a_1, ..., a_k$, real learning rates $(\alpha_i)_{i \in \{1,...,n\}}$, discount factor $\gamma$, initial $\epsilon$, decay factor $\delta$, set of risky states $\mathcal{R}$, risk-factor $\rho$

**Result:** $Q^{(N)}$-values as estimates for $Q$-values

1 $Q^{(0)} \leftarrow (0, ..., 0)$
2 $s_t \leftarrow s_0$
3 **for** *i=1,...,N* **do**
4     sample a random action $a$ $\epsilon$-greedy
5     update $\epsilon$ as $\epsilon \leftarrow \delta \cdot \epsilon$
6     sample next state $s_{t+1}$ and cost $C_{t+1}$ using the simulator for $\mathcal{M}$
7     **if** $s_{t+1} \in \mathcal{R}$ **then**
8         $C_{t+1} \leftarrow \rho \cdot C_{t+1}$
9     **end**
10     **for** $(s', a') \in \mathcal{S} \times \mathcal{A}$ **do**
11         **if** $s' = s_t$ & $a' = a$ **then**
12             $V(s_{t+1}) \leftarrow \max_{a_j = a_1, ..., a_k} Q^{(i-1)}(s_{t+1}, a_j)$
                $Q^{(i)}(s', a') \leftarrow$
                $(1 - \alpha_i) Q^{(i-1)}(s', a') + \alpha_i (C_{t+1} + \gamma V(s_{t+1}))$
13         **else**
14             $Q^{(i)}(s', a') \leftarrow Q^{(i-1)}(s', a')$
15         **end**
16     **end**
17     $s_t \leftarrow s_{t+1}$
18 **end**
19 **return** $Q^{(N)}$

## VII. Simulation Results

### A. Reference schemes

This section contains numerical results for the evaluation of the proposed threshold-based strategies as well as of $Q$-learning using risky states $Q$+RS for AoI, AoII, and QAoI including evaluations of the strategies' risk-sensitivity. We compare our results with three reference schemes. The first reference is a random strategy choosing to wait or to send both with a probability of $\lambda$ and independently of the current state. This rate corresponds to the probability of the arrival of a new packet. As a second reference, we use traditional risk-neutral tabular $Q$-learning (TQL). The third reference scheme is given by the TARQ scheme proposed in [21]. This strategy sends as long as the AoI at the sender is below a given threshold. As this threshold, here, we use the respective costwise optimal threshold for this strategy. We excluded the value iteration approach introduced in [15] from our comparison due to its inherent computational complexity. Without introducing a small limit for the AoI, the value iteration approach becomes impractical, making it less suitable for our study.

### B. Simulation setup

To simulate the system, we fix the parameters for transmission energy $\nu := 1$ and the channel's successful transmission probability $p := 0.9$. The weights in the cost function are set to $\alpha = 1$ and $\beta = 3$. These weights are chosen such that the costs arising from a transmission attempt are high enough that it is not always costwise reasonable for the sender to choose the "send" action. For AoI and QAoI, the default update arrival probability is set to $\lambda := 0.5$.
When simulating $\mathcal{M}_{AoII}$ for the AoII, the probability of remaining in the same state is set to $p_r := 0.5$ in the underlying Markov chain. The number of states in the underlying process is set to 10, such that we get $p_t = \frac{1-0.5}{9}$. These choices guarantee a sufficient dissimilarity between AoI and AoII. To also guarantee a dissimilarity between QAoI and AoI, we choose $q := 0.2$ as the probability for a time step to be a query time step. Each time step has a probability of 20% to be a query time step independently of all other time steps. For $Q$-learning-based strategies, we use $N := 10^5$ (resp. $N := 10^6$) time steps for learning and a discount factor $\gamma := 0.7$. As risk-factor, we choose $\rho = 2$ and as risk-threshold, we use $\zeta := 5$. For the AoII, we use $\zeta_{AoII} := 3$. Initially, $\epsilon = 0.9$, the decay factor is set to $\delta = 0.999$.

For the experiments, we take the average of 100 independent runs of the simulation. For the experiments used to find the best threshold-based strategies, we do not need any learning

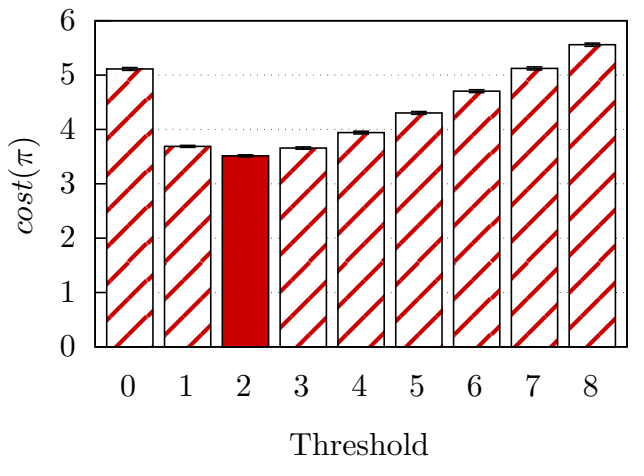

(a) Average $cost(\pi)$, which represents the weighted sum of AoI and energy.

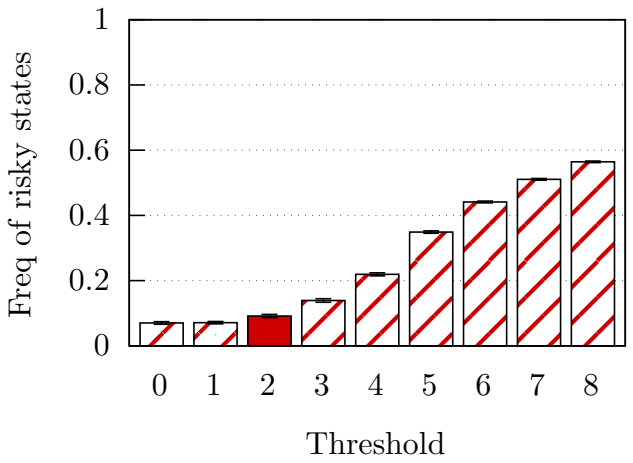

(b) Frequency of risky states, where risky states are states with $\text{AoI}_{Rx} \geq 5$.

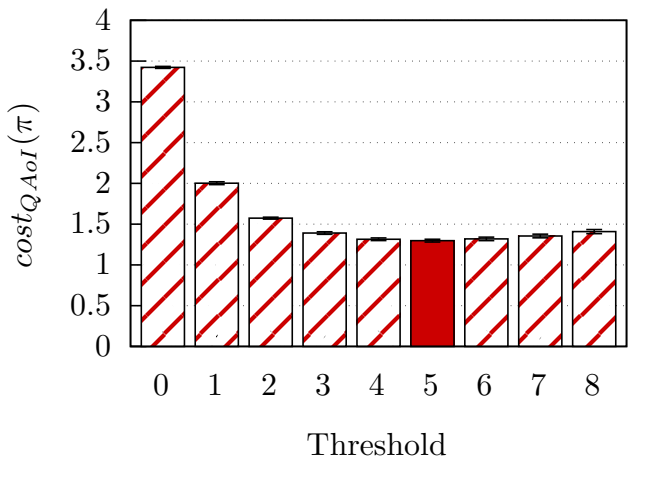

(c) Average $cost_{QAoI}(\pi)$, which represents the weighted sum of QAoI and energy.

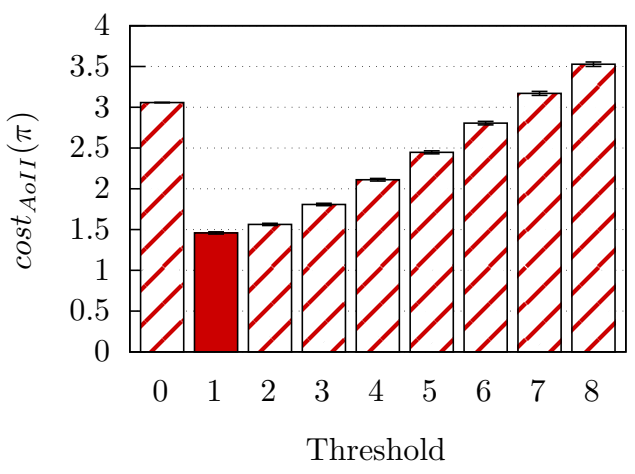

(d) Average $cost_{AoII}(\pi)$, which represents the weighted sum of AoII and energy.

Fig. 3: Numerical results for different threshold-based strategies. Bars representing the thresholds optimized for cost are highlighted in red.

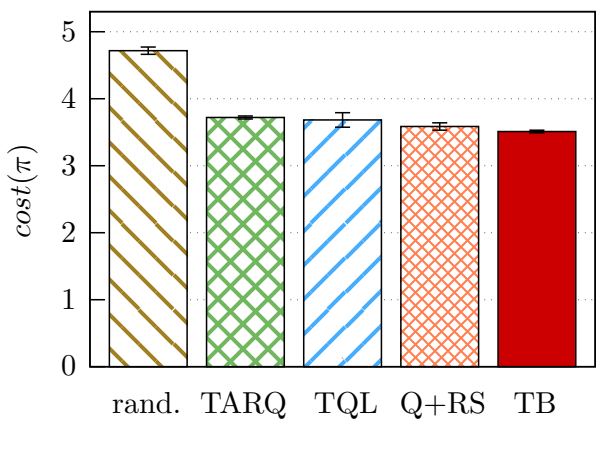

(a) Avg. AoI-cost, i.e. the weighted sum of AoI and energy.

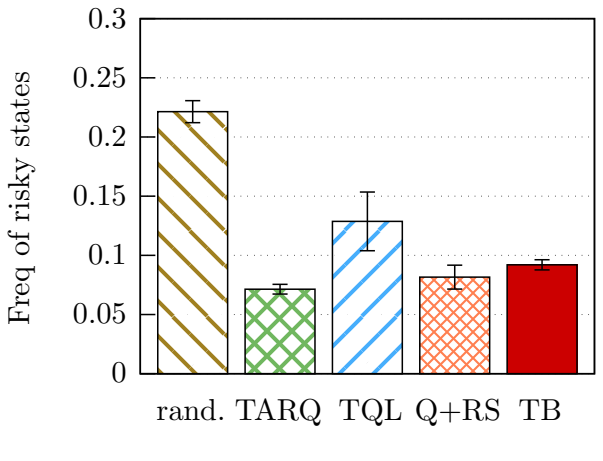

(b) Frequency of risky states for $\zeta = 5$.

Fig. 4: AoI-setting after $10^5$ learning time steps.

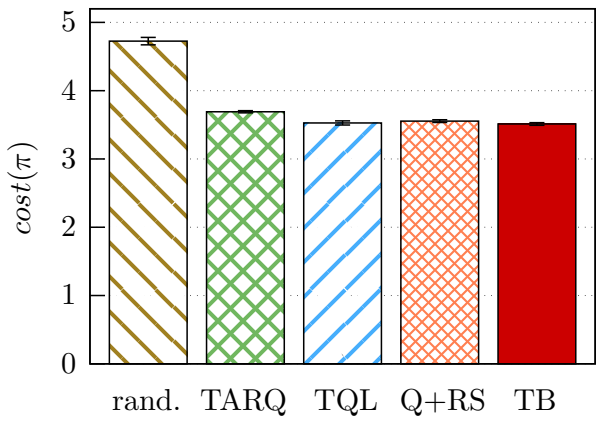

(a) Avg. AoI-cost, i.e. the weighted sum of AoI and energy.

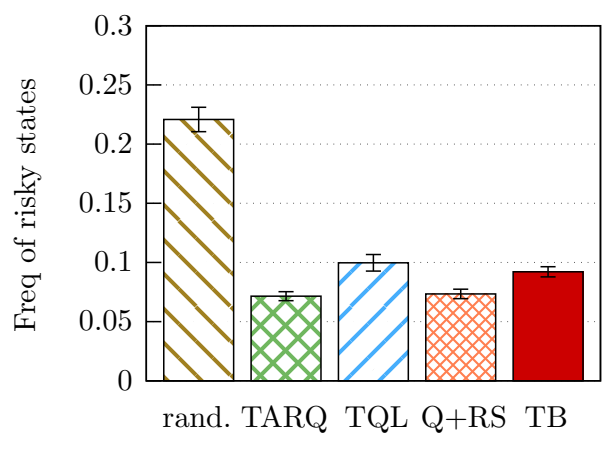

(b) Frequency of risky states for $\zeta = 5$.

Fig. 5: AoI-setting after $10^6$ learning time steps.

steps. In each run of the remaining experiments, we first train the $Q$-learning-based approaches. We then use the resulting learned strategies and compare them with the reference schemes. In each run, we use $10^4$ time steps per strategy for testing.

### C. Numerical results

The following results are organized to distinguish between risk and costs. Both measures were applied concurrently during the learning process. This enables a straightforward assessment of how each strategy addresses the metrics of cost and risk. All results include standard deviation error bars.

In Figure 3a, the empiric search for the optimal threshold is illustrated. Standard deviation error bars are indicated but small. From Lemma 1, we can derive that the costwise optimal threshold is 2. This is confirmed by Figure 3a. In Figure 3b, the corresponding frequencies for visiting risky states are shown. In case the probability of visiting risky states for TB-baseline is too high, it is reasonable to use a lower threshold.

In Figure 3c the same empiric search for the best threshold is shown when measuring the QAoI for $q = 0.2$. In this case, the costwise optimal threshold is 5. For AoII, we find the threshold 1 in Figure 3d. Note that this last result is empiric and not supported by a corresponding lemma as it is for the AoI and the QAoI.

Figure 4 shows the results for the AoI of our proposed strategies $Q$+RS and TB-baseline (short TB) compared to the reference schemes after $10^5$ learning time steps.

In Figure 4a, we show the average AoI-cost. If the system parameters $p$ and $\lambda$ are known, the optimal threshold-based strategy TB-baseline outperforms all other strategies. Otherwise TB is not applicable. In this case, when $p$ and $\lambda$ are unknown, the learning strategies offer a viable alternative. Both $Q$-learning-based strategies are able to perform close to TB-baseline in comparison to the random reference strategy. TQL generates average costs $4.9\%$ higher than that of the optimal threshold-based strategy. With this result, TQL matches the performance of the TARQ strategy from [21] already after $10^5$ learning time steps. The strategy derived from $Q$+RS generates costs only $2.1\%$ higher than that of the TB-baseline and has a $49\%$ lower standard deviation than TQL. Comparing the costs of our proposed strategies to TQL, $Q$+RS is reducing the average cost by $2.6\%$, while TB-baseline is reducing it by $4.7\%$. Compared to TARQ, $Q$+RS improves the average cost by $2.8\%$ and TB-baseline by $4.8\%$.

Figure 4b shows the average frequency of the appearance of *risky states* for our approaches and the reference strategies for the same simulations we used to measure the costs shown in Figure 4a. The strategy derived from $Q$+RS avoids those states actively and hence has a low frequency of $8.2\%$ compared to $22.1\%$ in the random case and $13.5\%$ for TQL. TARQ visits even less risky states than $Q$+RS but only in exchange for higher costs. A similar behavior can be reached for $Q$+RS by using a higher risk-factor $\rho$. TB-baseline visits risky states in $9.2\%$ of the time steps. Note that although TB-baseline was designed to minimize costs, it still is risk-sensitive in the sense that it visits a low number of risky states compared to TQL or the random strategy. To increase TB's risk-sensitivity further, it is possible to use a lower than the costwise optimal threshold. The frequency of risky states would decrease to $7.0\%$ for $n = 0$ and to to $7.1\%$ for $n = 1$ but only at the expense of higher risk-neutral costs (5.11 for $n = 0$ and 3.69 for $n = 1$ respectively instead of 3.51 for $n = 2$ and compared to 3.59 for $Q$+RS) We derive two crucial insights from these initial results. First, risk-sensitive strategies can effectively achieve both, lower risk and reduced costs simultaneously, as demonstrated by the case of $Q$+RS. Second, that TB-baseline is capable of translating its knowledge advantage over

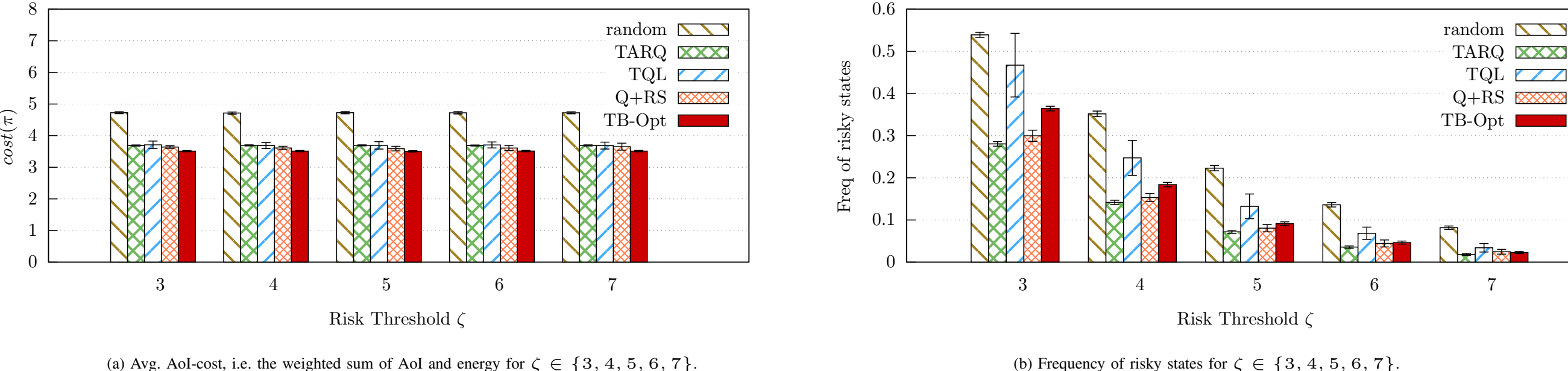

(a) Avg. AoI-cost, i.e. the weighted sum of AoI and energy for $\zeta \in \{3, 4, 5, 6, 7\}$.

(b) Frequency of risky states for $\zeta \in \{3, 4, 5, 6, 7\}$.

Fig. 6: AoI-setting after $10^5$ learning time steps for different risk thresholds $\zeta$.

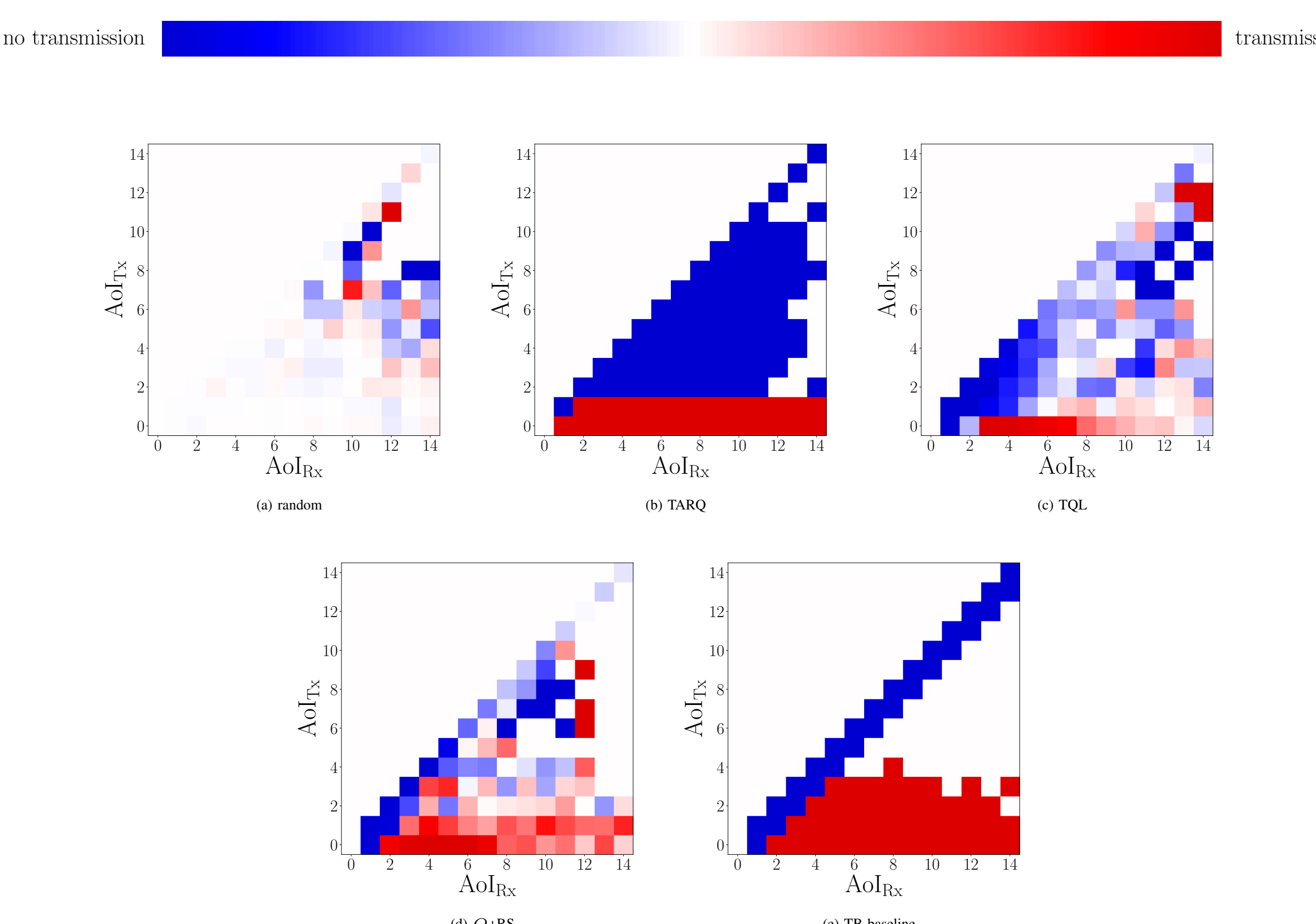

(a) random

(b) TARQ

(c) TQL

(d) $Q$+RS

(e) TB-baseline

Fig. 7: State-Action diagrams for the AoI-setting after $10^5$ learning time steps.

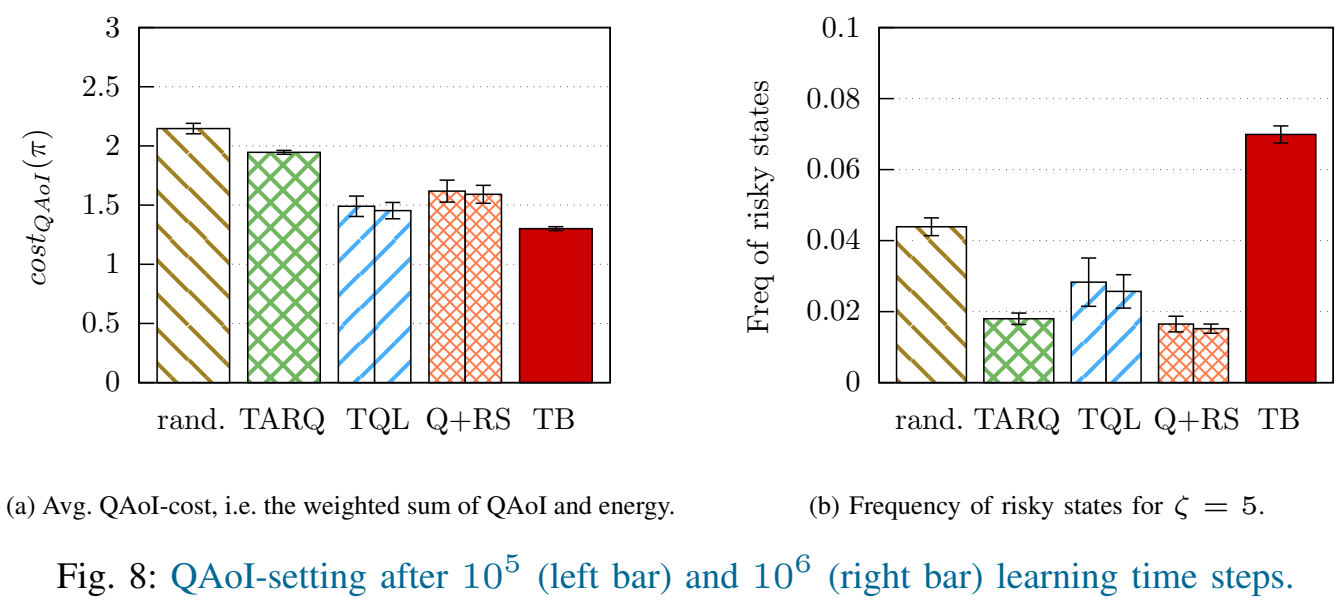

(a) Avg. QAoI-cost, i.e. the weighted sum of QAoI and energy.

(b) Frequency of risky states for $\zeta = 5$.

Fig. 8: QAoI-setting after $10^5$ (left bar) and $10^6$ (right bar) learning time steps.

system parameters directly into excellent performance in both measures.

We now compare the results after $10^5$ learning time steps in Figure 4 to the same strategies after $10^6$ learning time steps in Figure 5. Since the random strategy, TARQ, and TB-baseline do not depend on the learning phase, the results are very similar to those after $10^5$ time steps. The differences are due to the different outcomes of the random variables during the experiments. In Figure 5a, after $10^6$ learning time steps, $Q$+RS achieves costs of 3.55, close to that of TQL (3.53) and TB-baseline (3.51). TARQ generates higher costs of 3.69. Figure 5b shows that after $10^6$ learning time steps, $Q$+RS visits significantly less risky states than both TQL and TB-baseline. On average, $Q$+RS visits risky states in 7.3% of all time steps, which is close to TARQ (7.1), which trades risk-sensitivity for higher costs. At the same time, the frequency of risky states is still at 9.2% for TB-baseline and at 10.0% for TQL respectively. This means that although the costs are almost the same for both learning strategies, the frequency of visiting risky states is lowered by 26% by $Q$+RS compared to

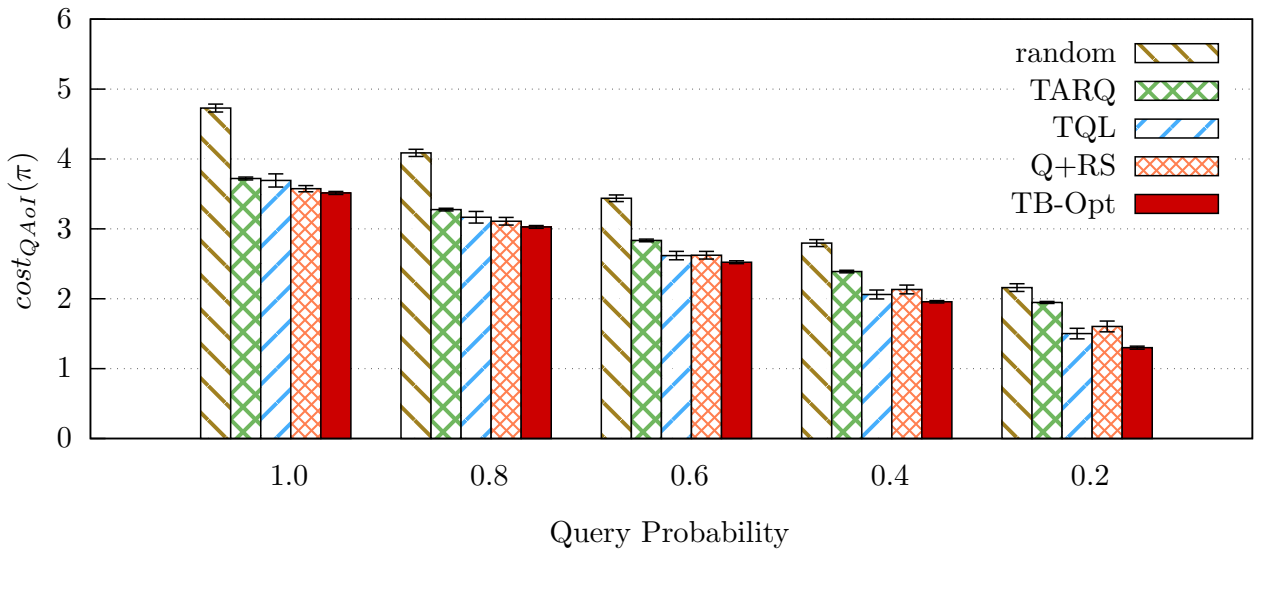

(a) Avg. QAoI-cost, i.e. the weighted sum of QAoI and energy.

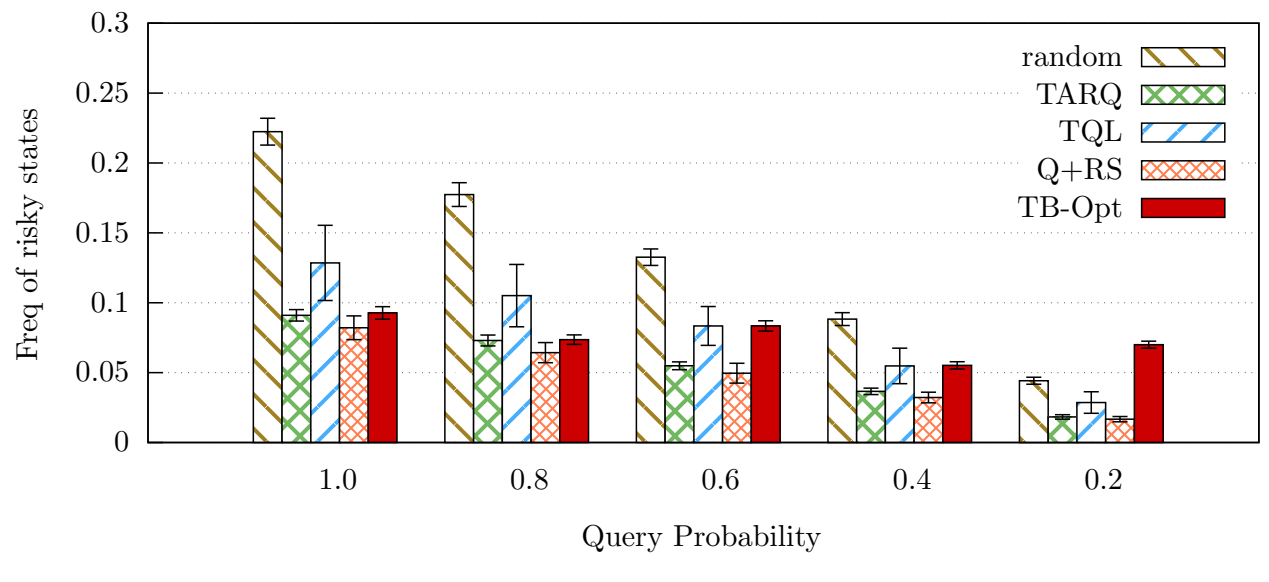

(b) Frequency of risky states for $\zeta = 5$.

Fig. 9: QAoI-setting after $10^5$ learning time steps for different query probabilities $q$.

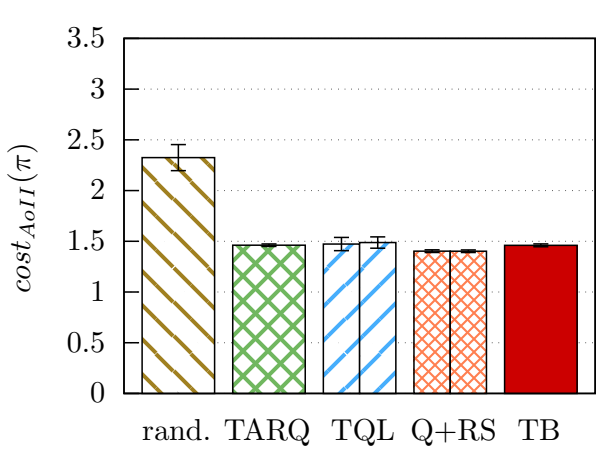

(a) Avg. AoII-cost, i.e. the weighted sum of AoII and energy.

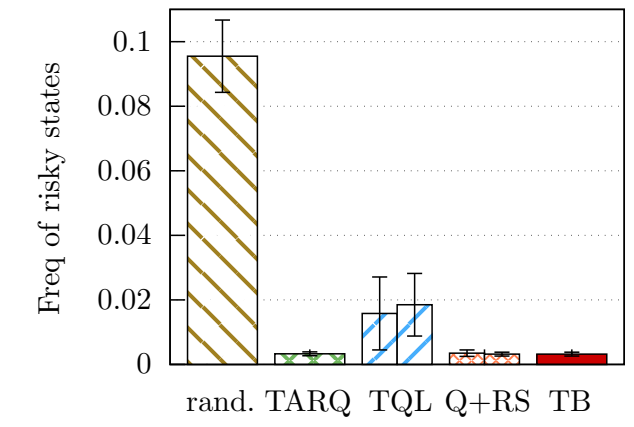

(b) Frequency of risky states for $\zeta_{AoII} = 3$.

Fig. 10: AoII-setting after $10^5$ (left bar) and $10^6$ (right bar) learning time steps.

TQL.

In Fig. 6, we compare the results for different risk thresholds $\zeta$. We deviate by at most 2 from the case $\zeta = 5$ and present results from $\zeta = 3$ to $\zeta = 7$. Notably, in Fig. 6a, the results of the risk-neutral strategies (random, TARQ, TQL, and TB-baseline) vary only slightly. Only $Q$+RS is effected by the change of $\zeta$. The effect of this change is larger than for the risk-neutral strategies but it does not effect the qualitative result that after $10^5$ learning time steps, $Q$+RS achieves better results than TQL and TARQ, while TB-baseline achieves the best results. In Fig. 6b, the frequency of risky states decreases as expected with increasing risk threshold for every strategy. As for $\zeta = 5$, TARQ visits a low number of risky states in exchange for higher costs compared to $Q$+RS. Note that by adjusting the risk-parameter $\rho$ for $Q$+RS, the risk-sensitivity can be increased. Clearly, $Q$+RS visits less risky states than the random strategy and TQL. For $\zeta = 7$, this effect vanishes as the frequency of risky states decreases and hence $Q$+RS converges to TQL.

To directly compare the behaviour of the strategies in different states, we present state-action diagrams in Fig. 7. On the horizontal axis, the $\mathrm{AoI}_{\mathrm{Rx},t}$ is denoted and on the vertical axis, the $\mathrm{AoI}_{\mathrm{Tx},t}$ is denoted. A blue square means that in the respective state, the strategy idled, while a red square means that in the respective state, the strategy transmits. If the behaviour of a strategy was not consistent for a given state, its tendency is displayed in a blue or red shade. To calculate this tendency, we divide the number of transmissions in this state by the total visits of this state. States that were not visited during the tests are displayed in white. For all strategies, states above the diagonal are never visited as $\mathrm{AoI}_{\mathrm{Rx},t}$ cannot be lower than$\mathrm{AoI}_{\mathrm{Tx},t}$. $(0, 0)$ is not visited, as $\mathrm{AoI}_{\mathrm{Rx},t}$ is at most 1. For the random strategy, frequently visited states in the lower left corner of the diagram are neither red nor blue, as the random strategy transmits with the same probability as it idles. States in the upper right corner are less frequently visited, such that by chance, the random strategy had a tendency to either transmit in this state or not. For TARQ, we clearly see the transmission threshold at $\mathrm{AoI}_{\mathrm{Tx},t} = 2$. Below this threshold, the strategy transmits. Above this threshold, it idles. Similarly, we can observe the threshold for TB-baseline. In this case, the difference between $\mathrm{AoI}_{\mathrm{Rx},t}$ and $\mathrm{AoI}_{\mathrm{Rx},t}$ is relevant, such that we observe a broad blue diagonal with states in which TB-baseline idles. In both cases, as well as for the learning strategies, some states on the right border of the diagram are never visited during the simulated testing phase. For the learning strategies, it is visible that their behaviour is less stable than for TARQ and TB-baseline. Also, it becomes clear that $Q$+RS tends to transmit more often than TQL, which matches our expectations.

Fig. 8 shows the results for the QAoI of the reference schemes, the threshold-based baseline derived from Lemma 4 and $Q$+RS. For the learning strategies, the left bar represents the average cost over $10^4$ test time steps after $10^5$ learning time steps. The right bar shows the average cost during a test of $10^4$ test time steps after $10^6$ learning time steps.

Figure 8a shows the average QAoI-costs $cost_{QAoI}(\pi)$. Both TQL and $Q$+RS are adjusted to minimize $cost_{QAoI}(\pi)$ instead of $cost(\pi)$ as described in Section VI. TB generates the lowest average cost (1.30), followed by TQL (1.49 after $10^5$ learning time steps and 1.45 after $10^6$ learning time steps), $Q$+RS (1.62 after $10^5$ learning time steps and 1.59 after $10^6$ learning time steps), TARQ (1.95) and Random (2.15). We can deduce that as in the AoI case, both learning strategies have mostly completed their learning after $10^5$ learning steps. However, in contrast to the AoI case, TARQ is no longer able to perform similar to the learning strategies.

The benefits of $Q$+RS for QAoI are clearly visible in Figure 8b. First, it is noticeable that TB has a high number of risky states (7.0%). This is because, for QAoI, TB sends in fewer time steps to save energy. For TARQ, we can observe the opposite, namely a low number of risky states (1.8%) and high average costs. We also note that both learning strategies improve, when comparing the results after $10^5$ and $10^6$ learning time steps respectively. However, $Q$+RS has advantages compared to TQL and even compared to the more costly TARQ strategy. After $10^6$ learning time steps, TQL exhibits a frequency of visiting risky states at $2.6\%$ of all time steps. In contrast, $Q$+RS significantly reduces these numbers to $1.5\%$, demonstrating a $78.6\%$ reduction compared to TB, $42.3\%$ improvement compared to TQL and $16.7\%$ compared to TARQ. Notably, unlike the random strategy or TARQ, $Q$+RS achieves these benefits without incurring drawbacks in average costs.

To verify the general picture of these results, we additionally compare the outcomes of the proposed algorithms for different query probabilities. Figure 9 shows the results for experiments, where the probability for a time step to be a query time step lies between $0.2$ and $1.0$. We use $10^5$ learning time steps for the learning strategies.

Note that for $0.2$, the results are the same as in Figure 8. If the probability for a query time step is $1.0$, the QAoI is no longer different from the AoI, which means that the results for a query probability of $1.0$ are the same as for AoI in Figure 4. Given the parameters and when measuring AoI (i.e. for query probability $1.0$), we already identified that the threshold-based strategy with a threshold of 2 is the most cost-effective strategy. For a query probability of $0.2$, we identified a threshold of 5 to be the most cost-efficient. In Figure 9a, we additionally find that the costwise best threshold-based strategies for the remaining query probabilities use the thresholds 3 (for $0.4$ and $0.6$) and 2 (for $0.8$). The smaller the query probability $q$ gets, the larger is TQL's costwise advantage over $Q$+RS. However, both learning strategies outperform TARQ for every $q$, already after $10^5$ learning time steps. TARQ still performs significantly better than the random strategy. This matches the results we observed for AoI and QAoI after $10^5$ time steps above.
In Figure 9b, we compare the frequency of risky states for different query probabilities. For different parameters, the optimal threshold varies. The optimal threshold is 2 for the query probabilities $1.0$ and $0.8$, which results in a declining frequency of risky states. It changes to 3 for $0.6$, which results in an increasing frequency of risky states. The same pattern is repeated as the threshold remains 3 for the query probability $0.4$ and increases again to 5 for the query probability of $0.2$. The frequency of visiting risky states is substantially declining for the random strategy and TARQ, as both strategies keep sending with the same rate. This is always at the expense of high average costs, especially for lower $q$. In comparison, the decline of this frequency is lower for TQL. $Q$+RS is either comparable to or better than the threshold-based baseline strategy for every tested query probability. $Q$+RS has the lowest frequency of risky states for every $q$ and achieves this without incurring as high costs as TARQ.

Figure 10 compares the results of $Q$+RS and the threshold-based approach with the reference schemes for the AoII. The threshold for the threshold-based approach was derived empirically as illustrated in Fig. 3d. Figure 10a shows that for the AoII, none of the learning strategies improves substantially in between $10^5$ and $10^6$ learning time steps in terms of cost. $Q$+RS generates lower costs than TQL after $10^5$ as well as after $10^6$ learning time steps. After $10^6$ learning time steps, $Q$+RS generates average costs of $1.4021$, which is $5.7\%$ less than the cost generated by TQL ($1.4876$), $4.1\%$ less than the cost generated by TARQ ($1.4615$), and $4.0\%$ less than the cost generated by the threshold-based strategy ($1.4602$).
The benefits of $Q$+RS become again visible in Figure 10b, where the frequency of risky states is shown for AoII. First note that the threshold for risky states is $\zeta_{AoII} = 3$ in this case. We observe that $Q$+RS drastically improves the frequency of visiting risky states when compared to TQL (improvement of $82.7\%$ after $10^6$ learning time steps). After $10^6$ learning time steps, it outperforms TARQ and achieves the same performance as the threshold-based baseline strategy, visiting risky states at a rate of only $0.32\%$.

## VIII. Conclusions

In this work, we develop risk-sensitive transmission strategies for a point-to-point wireless communication scenario. We measure risk using the novel concept of risky states. Our cost metrics combine age-based metrics with energy costs. Specifically, we consider three age-based metrics, namely, AoI, QAoI and AoII. For all three metrics, we propose a threshold-based strategy and use an optimization approach to determine the costwise optimal threshold. Here, the costs are defined as the weighted sum of the age-based metric and the transmission energy. By decreasing the optimal threshold, we reduce the frequency of visited risky states, leading to a risk-sensitive strategy. Taking into account that the threshold-based approach requires knowledge of the relevant probabilities used in the model, i.e., status update arrival probability or the probability of a successful transmission, we additionally propose a modified Q-learning algorithm, Q+RS, where we incorporate risk penalties to the cost function. In our simulations, we demonstrate that both of our proposed strategies outperform the reference schemes. While the threshold-based approaches surpass the reference schemes in terms of cost, Q+RS exhibits significant advantages in terms of risk. Interestingly, in many scenarios, Q+RS outperforms traditional Q-learning in terms of risk and cost.

## Appendix A: Proof of Lemma 1

Here, we provide the proof for Lemma 1 as stated in Section III above.

**Lemma 1.** *The average cost of the strategy $\pi_{TB}(n)$ is*

$$cost(\pi_{TB}(n)) = \frac{\alpha\lambda p\left(\sum\limits_{r=1}^{n} w_r\left(a(n) - \frac{r(r-1)}{2}\right) + \sum\limits_{r=n+1}^{\infty} w_r(r + a(1))\right) + \beta\lambda\nu}{\lambda(1-p) + 1 + p(1-\lambda) + \sum\limits_{r=1}^{n-1} w_r(n-r)}, \tag{28}$$

*where*

$$a(n) := \frac{n-2}{\lambda} + \frac{n-2}{p} + \frac{(n-2)(n-1)}{2} + \frac{1}{\lambda^2} + \frac{1}{\lambda p} + \frac{1}{p^2} \tag{29}$$

*and*

$$w_r := (1-\lambda)^{r-1}(1-p)^{r-1} - (1-\lambda)^r(1-p)^r. \tag{30}$$

*Proof.* To find $cost(\pi_{TB}(n))$, we consider the periods between two successful transmissions. These periods have an average period length of $l \in \mathbb{R}$, which is measured in time steps. $l$ is given in Proposition 5 after this proof. $l$ will be used to calculate both, the average energy-cost $AEC_n$ as well as the average AoI-cost $AAC_n$ of the strategy. Adding both will result in the average cost $cost(\pi_{TB}(n)) = AAC_n + AEC_n$. We start by calculating the average energy-cost $AEC_n$ of $\pi_{TB}(n)$. For that, we need to know the share of time steps in which $\pi_{TB}(n)$ decides to send. To find this share, we will first note that the average number $m$ of sending attempts per period is given by

$$m = p\sum_{i=0}^{\infty}(1-p)^i(i+1) = \frac{1}{p}. \tag{31}$$

The share of time steps in which $\pi_{TB}(n)$ decides to send is now calculated as $m \cdot l^{-1}$. We directly deduce that the average energy-cost $AEC_n$ of $\pi_{TB}(n)$ has to be

$$AEC_n = \frac{\beta \cdot \nu \cdot m}{l}. \tag{32}$$

It remains to find the average AoI-cost $AAC_n$. For a given period, we call the value of $\mathrm{AoI}_{\mathrm{Rx}}$ in this first time step of a period $r$. $r$ depends on the value of $\mathrm{AoI}_{\mathrm{Tx}}$ in the *last* time step of the previous period, as this is the age of the packet that was successfully sent to the receiver. If $r$ is larger than $n$, the sender will continue sending as soon as a new packet arrives at the transmitter. If $r$ is smaller than the transmission threshold $n$, the sender will not transmit until $\mathrm{AoI}_{\mathrm{Rx}}$ reaches $n$.

With $A_r$ defined as the average sum of all $\mathrm{AoI}_{\mathrm{Rx}}$ in a period with initial $\mathrm{AoI}_{\mathrm{Rx}} = r$, we get that

$$AAC_n = \frac{\alpha\sum_{r=1}^{\infty} w_r A_r}{l} = \frac{\alpha\left(\sum_{r=1}^{n} w_r A_r + \sum_{r=n+1}^{\infty} w_r A_r\right)}{l}, \tag{33}$$

where

$$w_r := (1-\lambda)^{r-1}(1-p)^{r-1} - (1-\lambda)^r(1-p)^r \tag{34}$$

is the probability that the initial value of $\mathrm{AoI}_{\mathrm{Rx}}$ in a given period is $r$.

Assuming that $r \leq n$, we get for $A_r$ that

$$\begin{aligned} A_r &= \lambda\sum_{j=0}^{\infty}(1-\lambda)^j p\sum_{i=0}^{\infty}(1-p)^i\overline{\mathrm{AoI}_{\mathrm{Rx},(n,i,j,r)}} \\ &= \left(a(n) - \frac{r(r-1)}{2}\right) \end{aligned} \tag{35}$$

where

$$\overline{\mathrm{AoI}_{\mathrm{Rx},(n,i,j,r)}} := \left(\frac{(n+i+j)(n+i+j+1)}{2} - \frac{r(r-1)}{2}\right) \tag{36}$$

and

$$\begin{aligned} a(n) :=& \frac{n-2}{\lambda} + \frac{n-2}{p} + \frac{(n-2)(n-1)}{2} \\ &+ \frac{1}{\lambda^2} + \frac{1}{\lambda p} + \frac{1}{p^2}. \end{aligned} \tag{37}$$

Assuming that $r > n$, we get that

$$A_r = r + a(1). \tag{38}$$

For $AAC_n$, this results in:

$$\begin{aligned} &AAC_n = \frac{\alpha}{l}\left(\sum_{r=1}^{n} w_r A_r + \sum_{r=n+1}^{\infty} w_r A_r\right) = \\ &\frac{\alpha}{l}\left(\sum_{r=1}^{n} w_r\left(a(n) - \frac{r(r-1)}{2}\right) + \sum_{r=n+1}^{\infty} w_r\left(w_r\left(r + a(1)\right)\right)\right) \end{aligned} \tag{39}$$

Combining average AoI-cost $AAC_n$, average energy-cost $AEC_n$ and $l$ from Proposition 5 results in the expression in the statement of Lemma 1:

$$\begin{aligned} &cost(\pi_{TB}(n)) = AAC_n + AEC_n = \\ &\frac{\alpha\lambda p\left(\sum\limits_{r=1}^{n} w_r(a(n) - \frac{r(r-1)}{2}) + \sum\limits_{r=n+1}^{\infty} w_r(r + a(1))\right) + \beta\lambda\nu}{\lambda(1-p) + 1 + p(1-\lambda) + \sum\limits_{r=1}^{n-1} w_r(n-r)}. \end{aligned} \tag{40}$$

□

In the proof above, we used $l$, which will be derived in the following proposition.

**Proposition 5.** *The average length $l$ in time steps of a period between two successful attempts to transmit is given by*

$$l = \frac{1-p}{p} + 1 + \frac{1-\lambda}{\lambda} + \sum_{r=1}^{n-1} w_r \cdot (n-r), \tag{41}$$

*where $w_r$ is defined as in Lemma 1.*

*Proof.* The average length $l$ of a period again depends on the value $r_0$ of $\mathrm{AoI}_{\mathrm{Rx}}$ at the period's first time step. If in a given period, $r_0$ is greater than or equal to the sending threshold $n$, the sender will decide to send as soon as a new status update arrives. This results in an average period length of

$$\begin{aligned} l_{r_0 \geq n} &= \lambda\sum_{j=0}^{\infty}(1-\lambda)^j p\sum_{i=0}^{\infty}(1-\lambda)^i(i+j+1) \\ &= \frac{1-p}{p} + 1 + \frac{1-\lambda}{\lambda}. \end{aligned} \tag{42}$$

Otherwise, the sender will wait for $(n - r_0)$ time steps before sending newly arrived updates:

$$l_{r_0 < n} = (n - r_0) + l_{r_0 \geq n}. \tag{43}$$

Including the relevant probabilities results in

$$\begin{aligned} l &= \sum_{r=1}^{n-1} w_r \cdot l_{r_0<n} + \sum_{r=n}^{\infty} w_r \cdot l_{r_0 \geq n} \\ &= \frac{1-p}{p} + 1 + \frac{1-\lambda}{\lambda} + \sum_{r=1}^{n-1} w_r \cdot (n-r). \end{aligned} \tag{44}$$

□

## Appendix B: Proof of Lemma 3

Here, we provide the proof for Lemma 3 as stated in Section III above.

**Lemma 3.** *For the strategy $\pi_{TB}(n)$, the frequency $f_k$ of an AoI at the receiver of $k > n$ is given by*

$$f_k = \frac{w_k + P_n \cdot (\sum_{r=1}^{n} w_r) + \sum_{r=n+1}^{k-1} w_r P_r}{l}, \tag{45}$$

*where*

$$w_r = (1-\lambda)^{r-1}(1-p)^{r-1} - (1-\lambda)^r(1-p)^r \tag{46}$$

*and*

$$P_r := 1 - p\lambda \sum_{j=0}^{k-r-1} \sum_{j=0}^{i} (1-\lambda)^j (1-p)^{i-j} \tag{47}$$

*and $l$ is as in Proposition 5.*

*Proof.* As in the proof of Lemma 1, we will use the concept of periods. A period ranges from one successful transmission to the next. The average length of a period is given by $l$ as found in Proposition 5. Note that in every period, the event $\mathrm{AoI_{Rx}} = k$ will appear at most once. Whether the event $\mathrm{AoI_{Rx}} = k$ appears depends on the first value $r_0$ of $\mathrm{AoI_{Rx}}$ in the respective period. We want to find the probability $P_r$ that the $\mathrm{AoI_{Rx}}$ will be equal to $k$ at some time step in a given period with $r_0 = r$. If $r > k$, $\mathrm{AoI_{Rx}}$ will not take the value $k$ in this period ($P_r = 0$). If $r = k$ in the first time step of the period, the event $\mathrm{AoI_{Rx}} = k$ appears in this period ($P_r = 1$). If $r < n$, the sender waits until $\mathrm{AoI_{Rx}} = n$, which means that the probability for an $\mathrm{AoI_{Rx}}$ of $k$ in periods with $r < n$ is the same as in periods, where $r_0 = n$ ($P_r = P_n$). Then, by using $w_r$ from the previous proof, we get that

$$f_k = \frac{1}{l}\sum_{r=1}^{\infty} w_r P_r = \frac{1}{l}(w_k + P_n \cdot (\sum_{r=1}^{n} w_r) + \sum_{r=n+1}^{k-1} w_r P_r). \tag{48}$$

We still need to find $P_{r_0}$ for $r_0 \in \{n, ..., k-1\}$. In a given period starting with an $\mathrm{AoI_{Rx}}$ of $r_0 = r \in \{n, ..., k-1\}$, $k$ will appear if and only if it takes at least $k - r$ time steps until the next successful transmission. We now find $\Sigma_r$, which is the sum of all the probabilities for faster successful transmissions. Subtracting $\Sigma_r$ from 1 results in $P_r$.

As $r \geq n$, the sender chooses to send immediately as soon as a new status update arrives. The necessity for a new status update results in a factor $\lambda$ in $\Sigma_r$. The transmission is successful with a probability of $p$, which is the second necessary factor for $\Sigma_r$. In the remaining $k - r - 1$ time steps, the sender will first wait for a new update, resulting in an additional factor $(1 - \lambda)$. As soon as a new update arrives, the remaining time steps have to consist of failing attempts to transmit, each time resulting in an additional factor $(1 - p)$. Adding all possible sequences of waiting and failing resulting in successful transmissions before $\mathrm{AoI_{Rx}}$ reaches $k$, we get

$$\Sigma_r = p\lambda \sum_{j=0}^{k-r_0-1} \sum_{j=0}^{i} (1-\lambda)^j (1-p)^{i-j}. \tag{49}$$

Subtracting $\Sigma_r$ from 1 results in $P_r$ as in Lemma 3. □

## Appendix C: Convexity of $cost(\pi_{TB}(n))$

Here, we provide a proof sketch for the convexity of $cost(\pi_{TB}(n))$ as it was stated in Section III above.

**Proposition 6.** *The function $n \to cost(\pi_{TB}(n))$ for $n \in \mathbb{N}_0$ is convex in the sense that:*

$$n \to cost(\pi_{TB}(n)) - cost(\pi_{TB}(n-1)) \tag{50}$$

*is increasing in $n$.*

*Proof Sketch.* The statement can be proven by showing for $n \in \{2, 3, ...\}$ that:

$$\begin{aligned}
& cost(\pi_{TB}(n+1)) - 2cost(\pi_{TB}(n)) + cost(\pi_{TB}(n-1)) \\
&= \frac{\alpha\lambda p\left(\sum_{r=1}^{n+1} w_r(a(n+1) - \frac{r(r-1)}{2}) + \sum_{r=n+2}^{\infty} w_r(r + a(1))\right) + \beta\nu\lambda}{\lambda + p - \lambda p + \lambda p \sum_{r=1}^{n} w_r \cdot (n - r + 1)} \\
&\quad - 2 \cdot \frac{\alpha\lambda p\left(\sum_{r=1}^{n} w_r(a(n) - \frac{r(r-1)}{2}) + \sum_{r=n+1}^{\infty} w_r(r + a(1))\right) + \beta\nu\lambda}{\lambda + p - \lambda p + \lambda p \sum_{r=1}^{n-1} w_r \cdot (n - r)} \\
&\quad + \frac{\alpha\lambda p\left(\sum_{r=1}^{n-1} w_r(a(n-1) - \frac{r(r-1)}{2}) + \sum_{r=n}^{\infty} w_r(r + a(1))\right) + \beta\nu\lambda}{\lambda + p - \lambda p + \lambda p \sum_{r=1}^{n-2} w_r \cdot (n - r - 1)} \geq 0.
\end{aligned} \tag{51}$$

For the sake of readability, we define:

$$\begin{aligned}
c &:= \lambda + p - \lambda p \\
s_1 &:= \lambda p \sum_{r=1}^{n-2} w_r \cdot (n - r - 1) \\
s_2 &:= \lambda p \sum_{r=1}^{n-1} w_r \cdot (n - r) \\
s_3 &:= \lambda p \sum_{r=1}^{n} w_r \cdot (n - r + 1) \\
b_1 &:= \alpha\lambda p \left(\sum_{r=1}^{n+1} w_r(a(n+1) - \frac{r(r-1)}{2}) + \sum_{r=n+2}^{\infty} w_r(r + a(1))\right) \\
b_2 &:= \alpha\lambda p \left(\sum_{r=1}^{n} w_r(a(n) - \frac{r(r-1)}{2}) + \sum_{r=n+1}^{\infty} w_r(r + a(1))\right) \\
b_3 &:= \alpha\lambda p \left(\sum_{r=1}^{n-1} w_r(a(n-1) - \frac{r(r-1)}{2}) + \sum_{r=n}^{\infty} w_r(r + a(1))\right)
\end{aligned}$$

The inequality (51) above can then be rewritten as:

$$\begin{aligned}
&(c + s_2)(c + s_3)(b_1 + \beta\nu\lambda) \\
&-2(c + s_1)(c + s_3)(b_2 + \beta\nu\lambda) \\
&+(c + s_1)(c + s_2)(b_3 + \beta\nu\lambda) \geq 0.
\end{aligned} \tag{52}$$

To show that (52) holds, we can divide the inequality into two parts:

$$(c + s_2)(c + s_3) - 2(c + s_1)(c + s_3) + (c + s_1)(c + s_2) \geq 0 \tag{53}$$

and

$$(c + s_2)(c + s_3)b_1 - 2(c + s_1)(c + s_3)b_2 + (c + s_1)(c + s_2)b_3 \geq 0. \tag{54}$$

If both, (53) and (54) hold, then (52) holds. In (53), we already dropped the positive factor $\beta\nu\lambda$. We show that Eq. (53) holds:

$$\begin{aligned}
&(c + s_2)(c + s_3) - 2(c + s_1)(c + s_3) + (c + s_1)(c + s_2) \\
&= c \cdot (2s_2 - s_1 - s_3) + s_1 s_2 - 2s_1 s_3 + s_2 s_3 \\
&= -c\lambda p w_n + s_1 s_2 - 2s_1 s_3 + s_2 s_3
\end{aligned}$$

$$\geq - c\lambda p w_n + \sum_{r=1}^{n-1} w_r \left( 2 \sum_{q=1}^{n} w_q + w_n (r-n) \right)$$

This lower bound is clearly increasing in $n$. Hence, to show that it is positive for $n \in \{2, 3, ...\}$, it suffices to show that it is positive for the smallest $n$, namely for $n = 2$:

$$\begin{aligned} & - c\lambda p w_2 + \sum_{r=1}^{1} w_r \left( 2 \sum_{q=1}^{2} w_q + w_2 (r-2) \right) \\ =& 6\lambda p + 3p^2 - p^3 + 3\lambda^2 - \lambda^3 - 9\lambda^2 p - 9\lambda p^2 + 7\lambda^2 p^2 + 2\lambda^3 p + 2\lambda p^3 \\ & + 2\lambda^2 p^3 + 2\lambda^3 p^2 - 6\lambda^3 p^3 + \lambda^4 p + \lambda p^4 - 3\lambda^4 p^2 \\ & - 3\lambda^2 p^4 + 3\lambda^4 p^3 + 3\lambda^3 p^4 - \lambda^4 p^4. \end{aligned}$$

This multivariate polynomial in $\lambda$ and $p$ is minimal and $0$ at $\lambda = p = 0$ on the set $\lambda, p \in (0, 1)^2$, showing that (53) holds. By equivalently showing (54), also (52) holds, proving that (50) increases in $n$.

## Appendix D: Proof of Lemma 4

Here, we provide the proof for Lemma 4 as stated in Section IV above.

**Lemma 4.** *Assume that $Q \subseteq \mathbb{N}$ contains every natural number independently with a probability of q. Measuring the QAoI for this set of query time steps $Q$, the average cost of the strategy $\pi_{TB}(n)$ is given by*

$$cost_{QAoI}(\pi_{TB}(n)) = \frac{q\alpha\lambda p \left( \sum\limits_{r=1}^{n} w_r \left( a(n) - \frac{r(r-1)}{2} \right) + \sum\limits_{r=n+1}^{\infty} w_r (r + a(1)) \right) + \beta\lambda\nu}{\lambda(1-p) + 1 + p(1-\lambda) + \sum\limits_{r=1}^{n-1} w_r (n-r)}, \tag{55}$$

*where $a(n)$ and $w_r$ are defined as in Lemma 1.*

*Proof.* The proof is the same as the proof for the AoI case in Lemma 1 apart from the fact that we have to include the probability $q$ when calculating the average AoI-cost $AAC$. We can hence replace Eq. (33) by:

$$AAC_n = \frac{q\alpha \left( \sum_{r=1}^{n} w_r A_r + \sum_{r=n+1}^{\infty} w_r A_r \right)}{l}. \tag{56}$$

Consequently, we exchange Eq. (39) by:

$$AAC_n = \frac{q\alpha}{l} \left( \sum_{r=1}^{n} w_r A_r + \sum_{r=n+1}^{\infty} w_r A_r \right) = \frac{q\alpha}{l} \left( \sum_{r=1}^{n} w_r \left( a(n) - \frac{r(r-1)}{2} \right) + \sum_{r=n+1}^{\infty} w_r \left( w_r \left( r + a(1) \right) \right) \right). \tag{57}$$

In Eq. (40), this results in:

$$cost(\pi_{TB}(n)) = AAC_n + AEC_n = \frac{q\alpha\lambda p \left( \sum\limits_{r=1}^{n} w_r (a(n) - \frac{r(r-1)}{2}) + \sum\limits_{r=n+1}^{\infty} w_r (r + a(1)) \right) + \beta\lambda\nu}{\lambda(1-p) + 1 + p(1-\lambda) + \sum\limits_{r=1}^{n-1} w_r (n-r)}. \tag{58}$$

□